\documentclass[aps,prd,twocolumn,showpacs,floatfix,preprintnumbers,amsfont,amsmath,amssymb,nofootinbib, superscriptaddress]{revtex4-2}

\usepackage[utf8]{inputenc}
\usepackage{amsmath}
\usepackage{physics}
\usepackage{comment}
\usepackage{bm}
\usepackage{siunitx}
\usepackage{graphicx}
\usepackage{natbib}
\usepackage[hidelinks]{hyperref}
\usepackage{mathtools}
\usepackage{xcolor}
\usepackage{ctable}
\usepackage[normalem]{ulem}
\usepackage{multirow}
\usepackage[caption=false]{subfig}
\usepackage{enumitem}

\newcommand{\rmd}{{\rm d}}
\newcommand{\diff}[2]{\frac{\rmd #1}{\rmd #2}}
\newcommand{\msun}{{\rm M_\odot}}
\newcommand{\chieff}{{\chi_{\rm eff}}}
\newcommand{\chidiff}{{\chi_{\rm diff}}}
\newcommand{\mones}{{m_{1\rm s}}}
\newcommand{\mtwos}{{m_{2\rm s}}}
\newcommand{\mbreak}{{m_{\rm break}}}
\newcommand{\mmax}{{m_{\rm max}}}
\newcommand{\mpeak}{{m_{\rm peak}}}
\newcommand{\pastro}{{p_{\rm astro}}}
\newcommand{\pastroi}{{p_{{\rm astro}, i}}}
\newcommand{\ntrig}{{N_{\rm trig}}}
\newcommand{\zetap}{{\zeta_{\rm pos}}}
\newcommand{\zetan}{{\zeta_{\rm neg}}}
\newcommand{\zetapeak}{{\zeta_{\rm peak}}}
\newcommand{\coloneq}{\vcentcolon=}
\newcommand{\chieffmin}{{\chi_{\rm eff}^{\rm min}}}
\newcommand{\xphm}{{\texttt{IMRPhenomXPHM}}}

\newcommand{\changed}[1]{#1}

\begin{document}

\title{Distribution of Effective Spins and Masses of Binary Black Holes\texorpdfstring{\\}{ }from the LIGO and Virgo O1--O3a Observing Runs}

\author{Javier Roulet}
\affiliation{\mbox{Department of Physics, Princeton University, Princeton, NJ 08540, USA}}
\author{Horng Sheng Chia}
\affiliation{\mbox{School of Natural Sciences, Institute for Advanced Study, Princeton, NJ 08540, USA}}
\author{Seth Olsen}
\affiliation{\mbox{Department of Physics, Princeton University, Princeton, NJ 08540, USA}}
\author{Liang Dai}
\affiliation{\mbox{Department of Physics, University of California, Berkeley, 366 LeConte Hall, Berkeley, CA 94720, USA}}
\author{Tejaswi Venumadhav}
\affiliation{\mbox{Department of Physics, University of California at Santa Barbara, Santa Barbara, CA 93106, USA}}
\affiliation{\mbox{International Centre for Theoretical Sciences, Tata Institute of Fundamental Research, Bangalore 560089, India}}
\author{Barak Zackay}
\affiliation{\mbox{Department of Particle Physics \& Astrophysics, Weizmann Institute of Science, Rehovot 76100, Israel}}
\author{Matias Zaldarriaga}
\affiliation{\mbox{School of Natural Sciences, Institute for Advanced Study, Princeton, NJ 08540, USA}}

\begin{abstract}
The distribution of effective spin $\chieff$, a parameter that encodes the degree of spin--orbit alignment in a binary system, has been widely regarded as a robust discriminator between the isolated and dynamical formation pathways for merging binary black holes. Until the recent release of the GWTC-2 catalog, such tests have yielded inconclusive results due to the small number of events with measurable nonzero spins. In this work, we study the $\chieff$ distribution of the binary black holes detected in the LIGO--Virgo O1--O3a observing runs. Our focus is on the degree to which the $\chieff$ distribution is symmetric about $\chieff = 0$ and whether the data provides support for a population of negative-$\chieff$ systems. 
We find that the $\chieff$ distribution is asymmetric at 95\% credibility, with an excess of aligned-spin binary systems ($\chieff>0$) over anti-aligned ones. Moreover, we find that there is no evidence for negative-$\chieff$ systems in the current population of binary black holes. Thus, based solely on the $\chieff$ distribution, dynamical formation is disfavored as being responsible for the entirety of the observed merging binary black holes, while isolated formation remains viable. We also study the mass distribution of the current binary black hole population, confirming that a single truncated power law distribution in the primary source-frame mass, $\mones$, fails to describe the observations. Instead, we find that the preferred models have a steep feature at $\mones \sim \SI{40}{\msun}$ consistent with a step and an extended, shallow tail to high masses.
\end{abstract}

\maketitle

\section{Introduction}
\label{sec:introduction}

The growing number of gravitational wave sources observed by the LIGO and Virgo detectors is leading to an improved picture of the astrophysical population of binary mergers. The recent release of the second Gravitational-Wave Transient Catalog, GWTC-2 \cite{GWTC2}, by the LIGO--Virgo Collaboration (LVC) has roughly tripled the sample size of observed binary black hole mergers \cite{GWTC-1, Nitz:2018imz, Venumadhav:2019tad, Zackay:2019tzo, Venumadhav:2019lyq, fishing, Nitz:2019hdf} and is starting to offer hints about the astrophysical origin of these binary systems \cite{GWTC2_pop, Wong2021, Zevin2021, Bouffanais2021}.

Indeed, the distribution of binary black hole parameters (e.g.\ masses, spins, redshift) is an observable that allows us to test models of formation pathways for these systems. Proposed scenarios include dynamical assembly and hardening of binary black holes in dense stellar environments, such as globular clusters \cite{zwart1999black, o2006binary, sadowski2008total, downing2010compact, downing2011compact, Samsing:2013kua, PhysRevLett.115.051101, rodriguez2016binary, askar2016mocca}, nuclear star clusters \cite{antonini2016merging, petrovich2017greatly}, and young stellar clusters \cite{ziosi2014dynamics, mapelli2016massive, banerjee2017stellar, chatterjee2017dynamical}; isolated evolution of a binary star in the galactic field, which undergoes either a common envelope phase \cite{nelemans2001gravitational, belczynski2002comprehensive, voss2003galactic, belczynski2007rarity, belczynski2008compact, dominik2013double, belczynski2014formation, mennekens2014massive, spera2015mass, eldridge2016bpass, stevenson2017formation, mapelli2017cosmic, giacobbo2017merging, mapelli2018cosmic, kruckow2018progenitors, giacobbo2018progenitors} or a chemically homogeneous evolution \cite{marchant2016new, de2016chemically, mandel2016merging}; and binary mergers prompted by interactions with a supermassive black hole \cite{antonini2012secular}, gas and stars in the accretion disk of an active galactic nucleus \cite{mckernan2012intermediate, stone2016assisted, bartos2017rapid}, or additional companions in higher-multiplicity systems \cite{antonini2014black, kimpson2016hierarchical, antonini2017binary, liu2018black, Hamers2015}.

Since the individual components of the dimensionless spin vectors $\bm{\chi_1}$ and $\bm{\chi_2}$ are hard to measure~\cite{Vitale:2014mka, Purrer:2015nkh, Vitale:2016avz} and their directions generally evolve with time due to precession~\cite{Apostolatos:1994mx, Kidder:1995zr}, a well-known effective aligned-spin parameter was introduced \cite{Racine:2008qv, Ajith:2009bn, Santamaria:2010yb}
\begin{equation}
    \chi_{\rm eff} \coloneq \frac{\bm{\chi_1} + q \hskip 1pt \bm{\chi_2}}{1+q} \cdot \bm{\hat{L}}, \label{eqn:chieff}
\end{equation}
where $\bm{\hat{L}}$ is the unit vector along the Newtonian orbital angular momentum of the binary, $q = m_2 / m_1 \leq 1$ is the mass ratio. The effective spin is motivated by the fact that it can be measured relatively precisely in the data, and is approximately conserved throughout the binary coalescence after orbit averaging~\cite{Racine:2008qv}.
No less important, two of the main broad classes of binary black hole formation channels make predictions about qualitative features of the effective spin distribution that are robust to model uncertainties. Dynamical formation channels in general predict that the spins and orbit should be isotropically distributed and uncorrelated with each other. In particular, this implies that for these systems the $\chieff$ distribution is symmetric about 0. Isolated formation channels instead predict correlations in the spins and orbit directions due to mass transfer episodes or tidal interactions between the component stars.
As a result, the isolated scenario predicts a distribution of $\chieff$ with little support at negative values. Within this channel, a small fraction of mergers with negative $\chieff$ could still possibly be explained by anisotropic supernova explosions at the black holes formation, which impart a natal kick that can change the plane of the orbit and thus the value of $\chieff$ \cite{Rodriguez2016, Gerosa2018}. However, if these kicks were strong enough to invert the direction of the orbit in a sizeable fraction of the cases, they would also unbind the binaries so frequently that the observed rates would be hard to explain \cite{belczynski1999, Callister2020}.

In this work we will study in detail the degree to which these two qualitative features of the $\chieff$ distribution, namely its symmetry about $0$ and support at negative values, hold for the observed sample. Both features become hard to test if black hole spins are small, which is predicted from stellar evolution models \cite{Fuller2019, Bavera2020} and is also the case of most observations. Indeed, until the recent release of the GWTC-2 catalog these simple but general tests were mostly inconclusive due to the small number of events with measurable nonzero $\chieff$ \cite{Farr2017, Farr2018, Roulet2019, Roulet2020}. Including events from the O3a observing run, \citet{GWTC2_pop} first reported evidence for both features in the population: they found the $\chieff$ distribution to have a positive mean and support at negative values. Together, these observations suggest that neither dynamical nor isolated formation channels alone can explain the entirety of the detections. Combining this information with the observed mass distribution, \citet{Zevin2021, Bouffanais2021} reached a similar conclusion, and applied a further layer of interpretation to constrain uncertain parameters of physical models of binary black hole formation.
Here, we instead constrain a phenomenological description of the binary black hole population, more akin to the analysis of \citet{GWTC2_pop}.

The mass distribution is another observable that can inform binary black hole formation channels, as well as physical processes of stellar evolution. Of special interest is the high-mass end of the mass distribution observable by LIGO--Virgo, $m \gtrsim \SI{40}{\msun}$. Due to the (pulsational) pair instability supernova process, black holes with mass between $\sim \SI{45}{\msun}$ and $\SI{135}{\msun}$ are not expected to form from stellar collapse (``upper mass gap'') \cite{Fowler1964, Barkat1967, Bond1984, Heger2003, Farmer2019}. A natural way to produce black holes in this mass range is through mergers of lighter black holes. In dense environments these so-called ``second-generation'' black holes can become paired and merge again, emitting an observable gravitational wave signal. This process is contingent on retention of the remnant black hole, so its efficiency depends on the interplay between the merger kick (a recoil of the remnant black hole due to asymmetric gravitational wave emission at merger) and the local escape velocity. The magnitude of the kicks is sensitive to the spins of the merging black holes, smaller spins usually yielding smaller kicks. In turn, different types of dense environments have different escape velocities, typical numbers being 10--\SI{e2}{\kilo\meter\,\second^{-1}} for globular clusters and up to $\sim \SI{e3}{\kilo\meter\,\second^{-1}}$ for nuclear clusters. Second generation mergers do not happen for binaries formed in isolation. Some alternative pathways to produce black holes in this mass range may involve accretion of gas \cite{Safarzadeh2020} or extreme values of the $\rm^{12}C(\alpha,\gamma)^{16}O$ nuclear cross section, which can shift the location of the mass gap \cite{Farmer2019}, see \cite{Gerosa2021} and references therein for a recent review. On the observational side, current interferometers are particularly sensitive to mergers in this high-mass region of parameter space, which makes it a promising discriminator \cite{Fishbach2017}. Indeed, some events were observed to have significant support for one or both component black holes in this mass range (e.g.\ GW190521, GW190602\_175927, GW190706\_222641, GW190519\_153544, GW190929\_012149 \cite{GWTC2}, GW170817A \cite{fishing}). While analyses prior to O3a found evidence for a cut-off in the mass distribution at $\sim \SI{40}{\msun}$ \cite{Fishbach2017, Wysocki2019, Roulet2019, Roulet2020}, this picture changed with the inclusion of O3a and models with more structure, including a tail at high mass, became favored \cite{GWTC2_pop}.
Here, we will also explore parametric models of the primary mass distribution in order to validate these results.

Our main findings are:
\begin{enumerate}
    \item The $\chieff$ distribution is inconsistent with being symmetric about zero at the 95\% credible level, with aligned-spin binary systems $(\chieff > 0)$ predominating over those with anti-aligned spins $(\chieff < 0)$. This result provides some evidence against the formation scenario in which the entire population of binary black holes has isotropically-distributed spins, as predicted if all merging binary black holes are formed dynamically in dense stellar environments;
    
    \item We find no evidence for negative $\chi_{\rm eff}$ in the population, in contrast to  \citet{GWTC2_pop}. We are able to reproduce their results, but find that the parametrized model they used in order to reach this conclusion is disfavored by the data and that the inferred presence of negative spins is contingent on this parametrization;
    
    \item We find that the primary-mass distribution steepens at $\sim\SI{40}{\msun}$ and then flattens, with an extended tail to high masses whose detailed shape is hard to constrain with current data.
\end{enumerate}

This paper is organized as follows: in Section~\ref{sec:data}, we describe the data investigated in this work, our sample selection criteria, and the parameter estimation method used to infer the source parameters of the binary black holes. In Section~\ref{sec:exploration}, we conduct a model-free exploration of the data, with a special focus on the empirical distribution of $\chieff$. In Section~\ref{sec:modelselection}, we describe our statistical methods for model selection and apply them to several parametrized models for the distributions of the effective spin and primary mass. We conclude in Section~\ref{sec:conclusions}. We provide details of the sample of events that we use in Appendix~\ref{app:gold_sample}.

\section{Data} \label{sec:data}

The data explored in this work consists of the binary black hole events reported in the LVC GWTC-1~\cite{GWTC-1} and GWTC-2~\cite{GWTC2} catalogs, and those identified in the independent IAS O1--O2 catalog~\cite{Venumadhav:2019tad, Zackay:2019tzo, Venumadhav:2019lyq, fishing}. Some of the events reported in the IAS O1--O2 catalog have been independently confirmed by \citet{Nitz:2018imz, Nitz:2019hdf}. Following the main analysis conducted by the LVC in their population study~\cite{GWTC2_pop}, we exclude GW190814~\cite{Abbott:2020khf} in this work as it is an outlier with respect to the rest of the observed population, and for ease of comparison between our results and the LVC's results (see Section~\ref{sec:modelselection}).
We do not include the recent 3-OGC catalog \cite{3OGC}, which was published as this work was being completed.
A summary of the events used in this work is provided in Appendix~\ref{app:gold_sample}.

Depending on how easily our models for astrophysical signals and detector noise can account for the properties of a given trigger, some detections are more statistically significant than the others. \citet{Roulet2020} provided a framework to take this into account when using triggers of arbitrary significance. However, in order to simplify the interpretation of the results shown in Section~\ref{sec:exploration}, we find it convenient to define a ``gold sample'' of events that are confidently astrophysical in origin. For a similar reason, we also exclude from the gold sample those events that happened when a detector exhibited non-Gaussian noise transients, which makes estimation of their parameters and significance more challenging.
We include an event in the gold sample if (i) it was identified by at least two search pipelines with a false-alarm rate $\rm FAR < \SI{0.1}{yr^{-1}}$; and (ii) none of the detectors exhibited non-Gaussian transient noise in its vicinity (see Appendix~\ref{app:gold_sample} for details). These criteria are neither explicitly dependent on nor expected to correlate significantly with the binary black hole intrinsic parameters; as such, our gold sample constitutes an unbiased representation of the distribution for the intrinsic parameters of detectable mergers. Indeed, as we shall see in Section~\ref{sec:modelselection}, our conclusions are not strongly affected by this choice of sample. Out of the total $55$ events considered in this work, $30$ are in the gold sample (see Appendix~\ref{app:gold_sample}).

We infer the source parameters of each binary system with the {\xphm} model, which describes the gravitational waves emitted by a quasi-circular binary black hole \cite{Pratten2020}. This model accounts for spin--orbit precession and the $(\ell, |m|)=\{(2, 2), (2, 1), (3, 3), (3, 2), (4, 4)\}$ harmonics of the gravitational radiation. We use the relative binning algorithm to evaluate the likelihood \cite{Zackay2018}, and \texttt{PyMultiNest}~\cite{Buchner2014} to sample the posterior distribution.
For the events identified near non-Gaussian transient noise (summarized in Ref.~\cite[table V]{GWTC2}), we do not make special mitigation efforts, though we verify that we obtain parameter estimation results that are similar to those reported by \citet{GWTC2}, who applied glitch subtraction algorithms before performing parameter estimations~\cite{Cornish2015, Littenberg2016, Cornish2021}.

For each event, we sample the posterior distribution using a prior that is uniform in detector-frame component masses, $\chieff$ and luminosity volume. For the remaining spin components, we adopt a uniform prior for the poorly-measured variable $\chidiff \coloneq (q \hskip 1pt \bm{\chi_{1}} - \bm{\chi_{2}} ) \cdot \bm{\hat{L}}  / (1+q)$, conditioned on $\chieff$ and enforcing the Kerr limit on the individual spin magnitudes, $\abs{\bm{\chi_1}}\leq 1 $ and $\abs{\bm{\chi_2}} \leq 1$. $\chieff$ and $\chidiff$ together determine the two spin components that are aligned with the orbital angular momentum,  $\chi_{1z}$ and $\chi_{2z}$. We then take the prior of the in-plane spin components of the black holes, $\chi_{ix}$ and $\chi_{iy}$ with $i=1, 2$, to be uniformly distributed in the disk $\chi_{ix}^2 + \chi_{iy}^2 \leq 1 - \chi_{iz}^2$.

Our parameter estimation results are broadly consistent with LVC's after accounting for the differences in spin priors, with two notable exceptions. Firstly, we find that the posterior distribution for GW151226~\cite{Abbott:2016nmj} significantly changes towards more unequal mass ratio, larger positive $\chieff$ and more misaligned primary spin when higher harmonics and precession are included in the parameter estimation~\cite{Chia2021}. Another remarkable event is GW190521, which was reported to have component source-frame masses $\mones = 85^{+21}_{-14}\,\msun$, $\mtwos = 66^{+17}_{-18}\,\msun$ by the LVC \cite{GW190521}. Using a different prior for the masses and distance, and allowing for a broader parameter range, \citet{Nitz2021} found a qualitatively different trimodal solution, with roughly similar total mass and peaks at $q \sim 1/2, 1/5,$ and $1/12$.
Instead, we find a bimodal solution which is approximately consistent with the first two of these modes \cite{Olsen2021}, similar to that reported in \cite{Estelles2021}.

\section{Model-free exploration}
\label{sec:exploration}
In this section, we carry out a model-free exploration of the data. Our emphasis is on the symmetry, or lack thereof, between positive and negative values of $\chieff$ in the observed $\chieff$ distribution. We also investigate if the data requires a distribution with support at negative values of $\chieff$. To ease the interpretation of the plots shown in this section, we shall restrict ourselves to the events identified in the gold sample (see Section \ref{sec:data}). We defer a model-dependent analysis of the data to Section~\ref{sec:modelselection}.

\subsection{Support for nonzero \texorpdfstring{$\chieff$}{effective spin}} \label{subsec:non-zero support}

\begin{figure*}[ht]
    \centering
    \includegraphics[width=\linewidth]{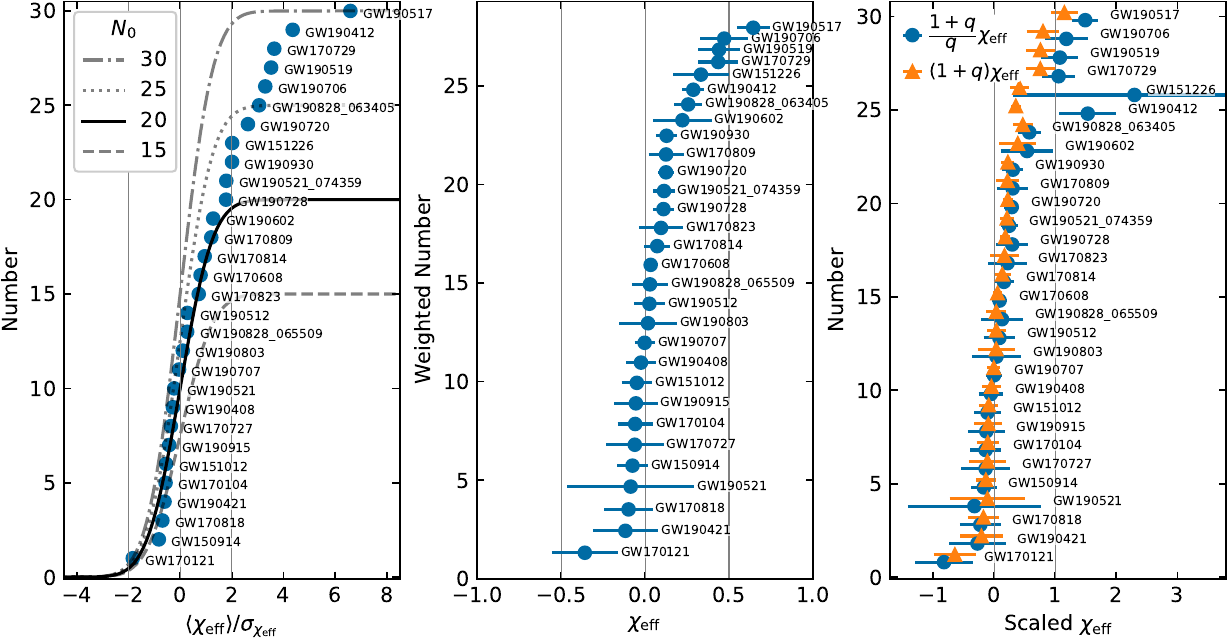}
    \caption{Empirical spin distributions of the events in the gold sample (see Appendix~\ref{app:gold_sample}). For each event, parameter estimation samples were obtained using a uniform prior in $\chieff$. To avoid clutter, event names were abbreviated when this did not cause ambiguity.%
    \textit{ Left panel:} Mean effective spin scaled by the standard deviation for each event's posterior. We see that about $N_0\approx 20$ events in the gold sample are consistent with noisy measurements of a $\chieff = 0$ subpopulation, but a tail in the positive $\chieff$ end of the distribution is clearly needed in order to accommodate the remaining $\approx 10$ events. Conversely, no such tail seems to be needed at the negative end.%
    \textit{ Middle panel:} $\chieff$ distribution, where markers and error bars indicate mean and standard deviation. In the cumulative, each event is weighted by the ratio of the event's sensitive volume to a similar event with zero spins in order to cancel spin selection effects. We see that there are several events with small but well-measured $\chieff>0$ for which spin selection effects are not important.%
    \textit{ Right panel:} ratio of observed $\chieff$ to its characteristic value if strong tides were acted on the secondary (blue circles) or primary (orange triangles) black hole progenitor. A number of events are inconsistent with any of these variables being 0 or 1, thereby excluding the strong-tide model as the only mechanism generating black hole spins. 
    }
    \label{fig:chieff}
\end{figure*}

We first test the simplest hypothesis that all binary black holes have $\chieff = 0$, with any apparent deviation away from zero arising due to measurement uncertainty. This test is motivated by the fact that, while the $\chieff$ measurements of some of the events have most of their support at $\chieff < 0$ (GW170121, GW150914, GW170818, GW190421\_213856, GW170104, GW151012, GW190915\_235702, GW170727, GW190521, GW190408\_181802), none of them confidently excludes $\chieff=0$. In the left panel of Fig.~\ref{fig:chieff}, we explore whether the observed scatter in the $\chieff$ distribution is consistent with noisy measurements of a $\chieff=0$ population. We plot the empirical distribution of the quantity $\langle\chieff\rangle / \sigma$, i.e.\ the mean $\chieff$ of each of the event's posterior samples divided by their standard deviation, and compare it with the cumulative of a standard Gaussian distribution with zero mean and amplitude $N_0$, where $N_0$ is the number of events in this distribution. Provided that the likelihood is approximately Gaussian as a function of $\chieff$, these distributions should match if the true $\chieff$ were 0. In particular, with the current number of observed events, we would not expect to find events that are more than $2\sigma$ away from $\chieff=0$. In the left panel of Fig.~\ref{fig:chieff}, we observe that although $N_0 \approx 20$ out of the 30 events in the gold sample are consistent with noisy measurements of a $\chieff=0$ distribution, there is an excess of about 10 events with $\chieff>0$ that cannot be explained by measurement uncertainty. On the other hand, no such tail seems to be needed in the $\chieff < 0$ interval.

\changed{
A phenomenon that will be important in the interpretation of this tail is the so-called ``orbital hangup'' \cite{Campanelli2006}, which entails that other parameters being equal, mergers with large and positive values of $\chieff$ have a louder emission of gravitational waves. This effect induces a selection bias, as events with $\chieff>0$ are detectable to larger distances. Under the hypothesis tested in this Section, however, all events have $\chieff=0$ and thus the bias is unimportant.\footnote{
Strictly speaking, this observational bias may in principle enter through the parameter estimation prior: in order to match the observed amplitude of a signal, higher-$\chieff$ solutions are located farther out in distance and thus have more phase space volume available. In other words, a flat prior for the astrophysical $\chieff$ distribution is implicitly skewed towards positive $\chieff$ values when conditioned on the strain amplitude measured at the detector. As a result, noisy measurements of a $\chieff=0$ distribution would be slightly biased towards $\chieff>0$. However, under the hypothesis that the true $\chieff=0$ this effect is small, as the measured $\chieff$ would be small.}}

\subsection{Symmetry of the \texorpdfstring{$\chieff$}{effective spin} distribution}

\changed{Due to the orbital-hangup effect,} the observed excess of $\chieff > 0$ events relative to those with $\chieff < 0$ in the left panel of Fig.~\ref{fig:chieff} does not immediately imply that the astrophysical $\chieff$ distribution is asymmetric about $\chieff = 0$. This effect leads to a selection bias favoring more observations of $\chieff > 0$ events, even if the astrophysical $\chieff$ distribution is symmetric \changed{about zero} \cite{Ng2018, Roulet2019}. The observed excess of $\chieff > 0$ events thus requires careful interpretation. In the middle panel of Fig.~\ref{fig:chieff}, we plot the empirical distribution of $\chieff$ correcting for this observational bias. The bias is computed as follows: for each event, we compute the weight factor
\begin{equation}
    w = \langle V_\text{no\,spin} / V \rangle,
\end{equation}
which is inversely proportional to the sensitive volume $V$ to the corresponding event. Here, we approximate $V$ of a source that has (detector-frame) intrinsic parameters $\theta_{\rm int}= \{ m_1, m_2, \chi_{1z}, \chi_{2z} \}$ as
\begin{equation}
    V(\theta_{\rm int}) \propto \rho_0^3(\theta_{\rm int}) ,
\end{equation}
with $\rho_0$ the single-detector signal-to-noise ratio (SNR) of an overhead, face-on source at a fiducial luminosity distance with a fiducial sensitivity. $V_\text{no\,spin}$ is defined similarly but with $\chi_{1z} = \chi_{2z} = 0$. For simplicity, we set the in-plane spin components to zero and neglect cosmological evolution throughout this computation.  We then average the ratio of these two volumes over the posterior distribution of each event in order to obtain the weight $w$. In the middle panel of Fig.~\ref{fig:chieff}, we see that many of the events that deviate most significantly away from $\chieff=0$ have small values of $\chieff$ and hence a small impact in the sensitive volume. In particular, GW190728\_064510, GW190521\_074359, GW190720\_000836, GW190930- \_133541, GW190828\_063405 and GW190412 are $\gtrsim2\sigma$ away from $\chieff=0$ and have relatively small values of $\chieff \sim 0.1$--$0.25$. The vertical spacing between events in this plot is given by the volume weight $w$ of the event: for the first four of these events $w$ is approximately $0.95$, and for the last two approximately $0.75$. Since these are small volume corrections, there is no compelling reason as to why the same number of corresponding events on the negative side of $\chieff$ should not be seen, for an astrophysical $\chieff$ distribution that is symmetric about zero.
There are also events that are more than $2\sigma$ away from zero $\chieff$ but with relatively large values of $\chieff \sim 0.5$ (GW170729, GW190519\_153544, GW190706\_222641 and GW190517\_055101). The selection effect for these events is more appreciable, $w \sim 0.5$, so it would be easier to miss similar events with the opposite sign of $\chieff$.
Altogether, the left and middle panels of Fig.~\ref{fig:chieff} hint that the empirical effective spin distribution is consistent with a distribution with no support for negative spins, but not so much with one symmetric about $\chieff=0$.

\subsection{Testing tidal models}

Finally, we explore whether the observed events with positive $\chieff$ can be explained by a simple model of tides acting on the progenitor of one of the component black holes. The simplest and most extreme model for tides assumes that tides sourced by the companion are either very efficient at spinning up the progenitor star or negligible depending on the orbital separation after a common envelope phase, because tidal torques are very sensitive to the orbital separation.
Then, a fraction of the component black holes would come from tidally-torqued progenitors and would have a large, aligned spin $\chi_z \approx 1$ \cite{Kushnir2016, Zaldarriaga2017}.
If, barring tides, natal black hole spins were small \cite{Fuller2019, Bavera2020}, the $\chieff$ distribution would have peaks at $\chieff = 0, q/(1+q), 1/(1+q), 1$ when the tides were inefficient, torqued the progenitor of the secondary black hole, torqued the progenitor of the primary, or torqued both, respectively. In the right panel of Fig.~\ref{fig:chieff} we show the empirical distribution of $\chieff$ rescaled by the value under the hypotheses that either the secondary or the primary black hole is maximally spinning and aligned with the orbit. We find that several of the events with well-measured nonzero spin do not seem to be well explained by this model (GW190728\_064510, GW190521\_074359, GW190720\_000836, GW170809, GW190930\_133541 and GW190828\_063405). This is in agreement with earlier findings that either a less extreme model of tidal torques (as argued in \cite{Qin2018, Bavera2020}) or a distribution of natal spins with some dispersion is needed in order to explain the observed spins with tides \cite{Roulet2020}.

\section{Model selection} \label{sec:modelselection}

In order to validate and quantify our findings in Section~\ref{sec:exploration}, in this section we perform selection of parametric models for the observed binary black hole population. We first provide a brief review of our statistical framework, and then constrain the parameters of several models for the astrophysical effective spin and primary mass distributions.

\subsection{Statistical framework}
\label{sec:framework}

Following \citet{Roulet2020}, we evaluate the likelihood $P(\{d_i\} \mid \lambda)$ of an observed set of triggers $\{d_i\}$, given a phenomenological population model $\lambda$ for the distributions of binary black hole source parameters, as:
\begin{multline} \label{eq:likelihood}
    P(\{d_i\} \mid \lambda)
    \propto e^{-N_a(\lambda)} \prod_{i=1}^{\ntrig} \bigg(\diff{N_a(\lambda)}{N_a(\lambda_0)}\bigg |_{d_i} \pastroi(\lambda_0) \\+ 1 - \pastroi(\lambda_0) \bigg).
\end{multline}
Here, $N_a(\lambda)$ is the expected number of triggers of astrophysical origin under the population model $\lambda$ (as opposed to detector noise), over a fixed and arbitrarily chosen detection threshold; $\rmd N_a(\lambda) / \rmd N_a(\lambda_0)|_{d_i}$ is the ratio of expected densities, in data space, of astrophysical triggers similar to that of the $i$th event between the population model $\lambda$ and a fixed, arbitrary reference model $\lambda_0$; and $\pastroi(\lambda_0)$ is the probability of astrophysical origin of the $i$th trigger under the reference population model.
The data space contains observable quantities that carry information about the astrophysical population, like measured detector strains and derived detection statistics.
All the quantities described above depend on the search pipeline used; in addition, $N_a(\lambda)$ and the set of triggers itself depend on the detection threshold chosen.
Three ingredients are required in order to estimate these quantities: a set of software injections labeled by whether they exceed the detection threshold, to quantify the sensitivity of the search; posterior samples characterizing the parameters of each individual event; and the set of $\{\pastroi(\lambda_0)\}$ encoding the events' significance \cite{Roulet2020}.

Equation~\eqref{eq:likelihood} naturally factors into the product of likelihoods from searches on disjoint datasets, such as different observing runs.
Since the full strain data from observing runs O1 and O2 are publicly accessible \cite{GWOSC}, for these data sets we base our analysis on our searches for binary black holes \cite{pipeline, BBH_O2, fishing}. The strain data for O3a has also recently been released, and analyzed by \citet{3OGC} when this work was close to completion; we do not include these results here. The LVC provides a set of software injections with FAR estimates from the search pipelines they used (cWB, GstLAL, PyCBC and PyCBC BBH) \cite{gwosc_url}, and the GWTC-2 catalog itself which reports $\{\pastroi\}$ for the latter 3 pipelines \cite{GWTC2}. For O3a we use these data products, which are adequate for computing the quantities in Eq.~\eqref{eq:likelihood} with the following caveat. Our method requires knowing $\{\pastroi(\lambda_0)\}$ under some specific astrophysical model, which was not explicited in the GWTC-2 release. We take two alternative approaches: (i) we conservatively consider only O3a events that are in the gold sample, so that all $\pastro=1$ under any model allowed by observations; or (ii) we consider the same O3a binary black hole mergers as in Ref.~\cite{GWTC2_pop}---i.e.\ with an inverse false-alarm rate $\rm IFAR > \SI{1}{yr}$ in any pipeline and excluding GW190814---taking the reported $\pastro$ at face value and assigning it to an arbitrary model $\lambda_0$ featuring a broad distribution in black hole parameters, described below.
We will refer to these two samples as GWTC-1 + IAS + Gold O3a and GWTC-1 + IAS + GWTC-2, respectively.
We will find that our conclusions are not strongly affected by the sample used.
We implement the sample choices by setting appropriate thresholds on the IFAR, which are reported for both events and injections in GWTC-2. The O3a injections do not report whether they fall near a glitch (one of the criteria of the gold sample), but these should be present in only a few percent of the events given the reported rate of $\sim \SI{1}{glitch/min}$ \cite{GWTC2}.

Following \cite{Roulet2020}, we adopt a fiducial population model $\lambda_0$ that is described by the following distribution function:
\begin{equation} \label{eq:f0}
    f(\mones, q, \chieff, D_L \mid \lambda'_0) \propto m_{1 \rm s}^{-\alpha_0} D_L^2,
\end{equation}
where $D_L$ is the luminosity distance and $\alpha_0 = 2.35$. We adopted the $\lambda'$  notation for the parameters that control the shape of the distribution, while the rate $R$ controls its normalization, i.e.\ $\lambda = (R, \lambda')$. The ranges of the parameters in Eq.~(\ref{eq:f0}) are taken to be $m_{1\rm min} < \mones < m_{1 \rm max}$ and $q_{\rm min} < q < 1$, where $m_{1 \rm min} = \SI{3}{M_\odot}$, $m_{1 \rm max} = \SI{120}{M_\odot}$ and $q_{\rm min} = 1/20$.

\subsection{Spin distribution}
\label{ssec:spin_distribution}
Motivated by Fig.~\ref{fig:chieff} and the discussion in Sections~\changed{\ref{sec:introduction} and} \ref{sec:exploration}, as well as Refs.~\cite{Farr2018, GWTC2_pop}, we will consider a phenomenological model for the effective spin distribution that allows us to explore the degree of symmetry of the distribution about $\chieff=0$. This model will also allow us to quantify the support for positive and negative values of $\chieff$ in the population.

\begin{figure}[t!]
    \centering
    \includegraphics[width=\linewidth]{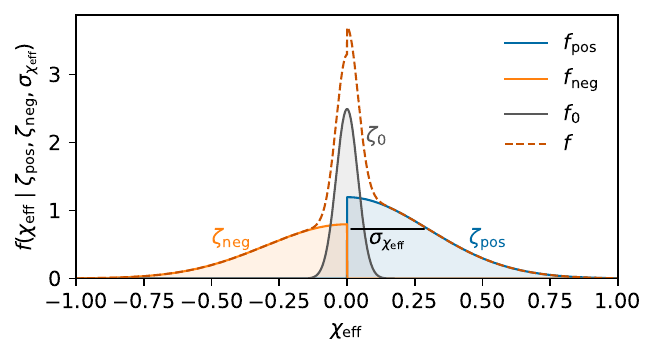}
    \caption{Sketch of the functional form Eq.~(\ref{eq:pos_vs_neg_model}), which we use to parametrize the $\chieff$ distribution as the sum of three subpopulations. These subpopulations have positive, negative or zero effective spins, with each described by truncated Gaussians that peak at $\chieff = 0$. We use three independent shape parameters: the width of the positive and negative distributions, $\sigma_{\chieff}$, which are constrained to be equal, and the three branching ratios $\zeta_j$ which sum to unity. For technical reasons, we fix the width of the $\chieff\approx0$ subpopulation to have a small but non-vanishing dispersion of $\sigma_0 = 0.04$.}
    \label{fig:sketch}
\end{figure}

Firstly, we model the effective spin distribution as a mixture of three subpopulations with negative, zero, and positive $\chieff$:
\begin{equation} \label{eq:pos_vs_neg_model}
    \begin{split}
        f_{\chieff}(\chieff \mid \zetap, \zetan, \sigma_\chieff)
        & = \zeta_0 \hskip 1pt \mathcal N(\chieff; \sigma_0 = 0.04) \\
        & + \zetan \hskip 1pt \mathcal N_{<0}(\chieff; \sigma_\chieff) \\
        & + \zetap \hskip 1pt \mathcal N_{>0}(\chieff; \sigma_\chieff) \, .
    \end{split}
\end{equation}
Here, the various parameters $\zeta_j \in [0, 1]$ are the branching ratios for each subpopulation, constrained to have unit sum; $\mathcal N(x; \sigma)$ is the normal distribution with zero mean, dispersion $\sigma$, truncated at $x = \pm 1$; $\mathcal N_{<0}(x; \sigma)$ is a similar normal distribution but truncated at $x=-1$ and $x=0$, while $\mathcal N_{>0}(x; \sigma)$ is truncated at $x=0$ and $x=1$. The functional form Eq.~(\ref{eq:pos_vs_neg_model}) is sketched in Fig.~\ref{fig:sketch} for a particular choice of parameters. Note that we have enforced the dispersion parameters of the positive and negative subpopulations to be equal, such that setting $\zetap = \zetan$ yields a $\chieff$ distribution \changed{symmetric about zero}. For the $\chieff \approx 0$ subpopulation, we adopt a small (relative to typical measurement uncertainties) but nonvanishing dispersion $\sigma_0 = 0.04$ in order to ensure that the reweighting procedure used in our algorithm is well behaved \cite{Roulet2020}. 

In this Section we will only vary the effective spin distribution, while the remaining spin components are assumed to follow the parameter estimation prior described in Section~\ref{sec:data}.
For the other binary black hole parameters, we will assume the following fixed distribution :
\begin{equation} \label{eqn:distribution_otherparams}
    f(\chieff, \mones, q, D_L)
    = f_\chieff(\chieff) f_\mones(\mones) f_q(q) f_{D_L}(D_L).
\end{equation}
Following \citet{GWTC2_pop} we adopt a broken power-law distribution for the primary mass:
\begin{equation} \label{eq:broken_power_law}
    f_\mones(\mones) \propto \begin{cases}
        0, &\mones < \SI{5}{M_\odot}\\
        \left(\dfrac\mones\mbreak\right)^{-\alpha_1},
            &\SI{5}{M_\odot} < \mones < \mbreak\\
        \left(\dfrac\mones\mbreak\right)^{-\alpha_2},
            &\mbreak < \mones,
    \end{cases}
\end{equation}
with $\alpha_1 = 1.6$, $\alpha_2 = 5.6$, $\mbreak = \SI{40}{M_\odot}$.
For simplicity, we adopt a mass-ratio distribution that is uniform in $1/20<q<1$ and take the distance distribution to be uniform in comoving volume-time.

\begin{figure}[b!]
    \centering
    \includegraphics[width=\linewidth]{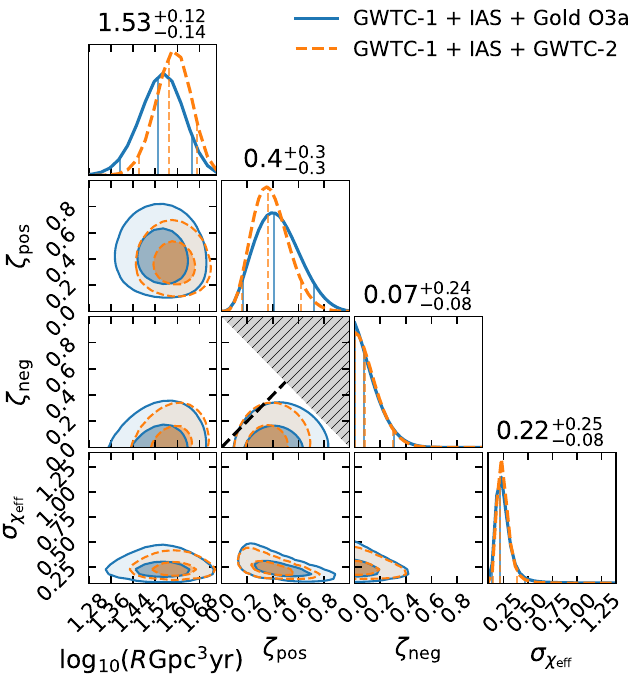}
    \caption{Constraints on the model parameters of the population model Eq.~\eqref{eq:pos_vs_neg_model}. We see that a $\chieff$ distribution \changed{symmetric about 0} (black dashed line), with $\zetap = \zetan$, is disfavored by the data. In addition, the population is consistent with having no negative-spin subpopulation. The two-dimensional contours enclose the 50\% and 90\% credible regions. Parameter values (median and 90\% confidence level) are reported for the GWTC-1 + IAS + Gold O3a sample.}
    \label{fig:positive_vs_negative_chieff}
\end{figure}

We use the likelihood in Eq.~\eqref{eq:likelihood} to obtain a posterior distribution for the population parameters, by adopting a Jeffreys prior for the overall merger rate $\pi(R \mid \lambda') \propto \sqrt{N_a(R, \lambda')}/R$; recall that $\lambda'$ are the shape parameters $(\zetap, \zetan, \sigma_\chieff)$. For these we adopt a uniform prior $\pi(\lambda') = \rm const$. This prior is invariant to the choice of which two out of the three branching ratios are used to parametrize the distribution.

We show our constraints on the parameters of this model in Fig.~\ref{fig:positive_vs_negative_chieff}, for the two samples used. We find two remarkable results: first, 95\% of the posterior lies at $\zetap>\zetan$ and a \changed{$\chieff$ distribution symmetric about zero ($\zetap = \zetan$, dashed line)} is disfavored; second, the population is consistent with $\zetan=0$, i.e.\ no spins anti-aligned with the binary orbit. These conclusions do not depend on which of the two event samples are considered. 

We quantify these statements using the Bayesian evidence and maximum likelihood as model scores: we report in Table~\ref{tab:scores} the scores achieved by the following models: a $\chieff$ distribution \changed{symmetric about zero} given by Eq.~\eqref{eq:pos_vs_neg_model} with $\zetap = \zetan$, a positive $\chieff$ distribution setting $\zetan = 0$, and the full mixture.
The symmetric $\chieff$ model is representative of a scenario completely dominated by dynamical formation in clusters, while the positive $\chieff$ model represents a case dominated by isolated binaries---with the caveat that in this channel there exist mechanisms to achieve some spin--orbit misalignment, e.g.\ supernova kicks.

The first result that positive $\chieff$ predominates over negative is in general agreement with the analysis of \citet{GWTC2_pop}. Indeed, parametrizing the $\chieff$ distribution with a Gaussian, they find that  a positive mean is preferred; likewise, they favor spin orientation distributions with at least some degree of anisotropy.

\begin{table}[t!]
    \centering
    \begin{tabular}{lll}
    \toprule
    {} & $\Delta \,\max \, \ln\, L$ &    $\Delta \,\ln \,Z$ \\
    \midrule
    Symmetric $\chieff$            &                        $0$ &                   $0$ \\
    Positive $\chieff$             &        $2.1_{-0.4}^{+0.5}$ &   $1.6_{-0.3}^{+0.5}$ \\
    Positive/Negative mixture $\chieff$ &        $2.1_{-0.4}^{+0.5}$ &   $1.4_{-0.2}^{+0.4}$ \\
    Gaussian $\chieff$             &        $0.2_{-0.6}^{+0.7}$ &  $-0.2_{-0.8}^{+0.6}$ \\
    \bottomrule
    \end{tabular}
    \caption{Scores for models of the $\chieff$ distribution. Difference in the maximum log likelihood and log evidence relative to the $\chieff$ model \changed{symmetric about zero}, $\zetap = \zetan$. Error bars indicate the 90\% confidence level and account for stochastic errors due to the finite number of injections and parameter samples used, and are estimated with 100 bootstrap realizations of the analysis similarly to \cite{Roulet2020}.}
    \label{tab:scores}
\end{table}

\begin{figure}
    \centering
    \includegraphics[width=\linewidth]{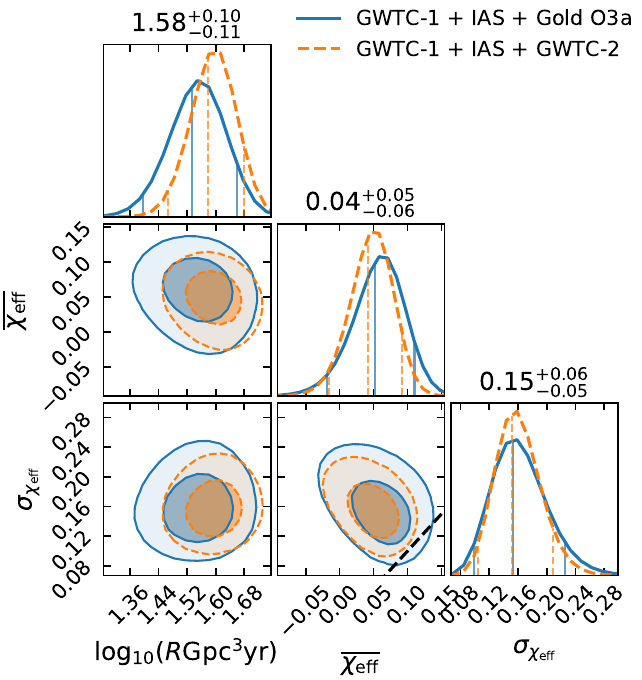}
    \caption{Constraints assuming a Gaussian model for the $\chieff$ distribution. The black dashed line corresponds to $\sigma_\chieff = \overline\chieff$, models above the line have sizable support at negative $\chieff$. Thus, contrary to Fig.~\ref{fig:positive_vs_negative_chieff}, under this model one would conclude that negative $\chieff$ are present in the population.}
    \label{fig:gaussian_chieff}
\end{figure}

On the other hand, our second finding that there is yet no evidence for negative $\chieff$ in the population is in contrast with the results of \citet{GWTC2_pop}, who found that all Gaussian fits to the observed $\chieff$ distribution had a sizable support at negative $\chieff$.
We suggest that their result is contingent on the assumed parametrization of the population as a Gaussian distribution, while our parametrization has more freedom to accommodate features near $\chieff=0$. In particular, the maximum likelihood solution has parameters $(\zetap, \zetan, \zeta_0, \sigma_\chieff) = (0.45, 0.00, 0.55, 0.23)$, featuring a sharp peak at $\chieff\approx0$, a rapid decline at negative $\chieff$ and an extended tail at positive $\chieff$ which are hard to capture with a single Gaussian. To test this hypothesis we try a similar Gaussian model for the $\chieff$ distribution, shown in Fig.~\ref{fig:gaussian_chieff}. With this model, we indeed find good quantitative agreement with \citet[figure~11]{GWTC2_pop} and would recover their same conclusions: we find that models without support at negative $\chieff$ ($\sigma_\chieff \ll \overline\chieff$) are excluded.
In Table~\ref{tab:scores} we see that the Gaussian model performs worse than other models we tried, in particular the model restricted to positive $\chieff$.
\citet{GWTC2_pop} did consider the possibility that their finding of negative spins could be driven by the Gaussian parametrization. Indeed, in \citet[figure~27]{GWTC2_pop} they show that adding a free parameter $\chieffmin$ below which the Gaussian is truncated, they exclude $\chieffmin\geq 0$ at 99\% credibility and find that small negative values $-0.2 \lesssim \chieffmin \lesssim 0$ are preferred. We interpret that the large number of events at $\chieff\approx0$ drives the exclusion of positive $\chieffmin$, furthermore, the fact that small negative values of $\chieffmin$ are preferred over large negative values indicates that the Gaussian model $\chieffmin = -1$, which motivated the claim of existence of negative $\chieff$ systems, does not fit well the observed population. We conclude that, while it is certainly possible that there are negative $\chieff$ systems in the population, there is not enough evidence for them yet.

Within isolated formation channels, the fraction of negative $\chieff$ systems $\zetan$ is an indicator of typical natal (supernova) kick velocities, larger kicks generally giving larger $\zetan$. \citet[figure 6]{Gerosa2018} find that measurements of $\zetan$ to a precision better than $0.1$ would start putting meaningful constraints on kick velocities. Our current bound $\zetan \lesssim 0.3$ is compatible with even extreme kicks, but with a factor of few more detections this would be a promising source of information.

We point out that the GWTC-1 + IAS + GWTC-2 sample differs from that of the analysis in \citet{GWTC2_pop} in that it includes events in the IAS catalog. However, having included these events only weakens our conclusions due to the presence of GW170121, the confident detection with the most support for negative $\chieff$ in the sample.

\subsection{Mass distribution}

We now turn to the distribution of merging binary black hole masses. Using data from the first two observing runs, several past studies have identified that the primary mass distribution was well described by a power-law truncated at $\mmax \approx \SI{40}{M_\odot}$ \cite{Fishbach2017, Wysocki2019, Roulet2019, GWTC1_pop, Roulet2020}. The third observing run revealed that the mass distribution has a tail that extends to higher masses, and that models with more features, e.g.\ a broken power-law, were favored. One diagnostic that a single truncated power-law did not fit the O3a data was that its inferred parameter values experienced a large shift when including the new events, in particular, $\mmax$ was found to increase from $40.8^{+11.8}_{-4.4}\,\rm M_\odot$ to $78.5^{+14.1}_{-9.4}\,\rm M_\odot$ \cite{GWTC2_pop}.

As this development evidenced, one has to bear in mind that with a finite number of events one cannot probe the tail of the distribution arbitrarily far out. Thus, constraints obtained on the population are to be interpreted as a characterization of the bulk of the distribution, up to a quantile that depends on the number of events: with $\ntrig$ triggers, a fraction $\sim\mathcal O(1 / \ntrig)$ of the distribution cannot be probed; with the present sample this is at the few-percent level.
At this point we introduce a feature in our analysis that makes this notion explicit: we add to the model a second subpopulation of astrophysical triggers that come from a broad parameter distribution $\lambda_0'$ accounting for a small fraction $\epsilon$ of the total rate:
\begin{equation} \label{eq:wildcard}
    \diff{N_a}{\theta}(\theta \mid \lambda, \lambda_0', \epsilon)
    = R \big[(1-\epsilon) f(\theta \mid \lambda')
             + \epsilon f(\theta \mid \lambda_0')\big];
\end{equation}
for $\epsilon = 0$ we recover the previous analysis. Recall that we call the distribution shape parameters $\lambda'$, so that $\lambda = (R, \lambda')$. For simplicity, we will fix the parameter $\epsilon = 0.05$. This change makes little difference for events that are well described by the population model $\lambda$, but since the broad subpopulation can accommodate any of its outliers, the model $\lambda$ is no longer forced to explain all the observations. A practical advantage of this is that we get a sensitive diagnostic that some specific events may be poorly accommodated by the (ultimately arbitrary) parametrizations we chose, if they get classified with high confidence as belonging to the other subpopulation $\lambda_0'$---evidencing that a model with more freedom is needed to explain all events. We also construct a simple goodness-of-fit test for the $\lambda$ model based on the Bayes factor between a model with $\epsilon = 0$ or a small fixed value $\epsilon=0.05$. If the $\epsilon=0$ model is already a good description of all observed events, adding a broad subpopulation should not increase the evidence significantly.

\begin{figure*}
    \centering
    \subfloat[Truncated power law model]{
    \includegraphics[width=.333\linewidth]{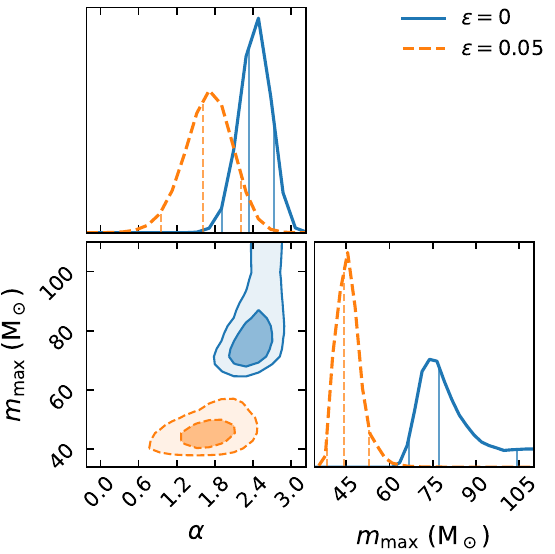}\label{fig:truncated}}
    \subfloat[Broken power law model]{
    \includegraphics[width=.333\linewidth]{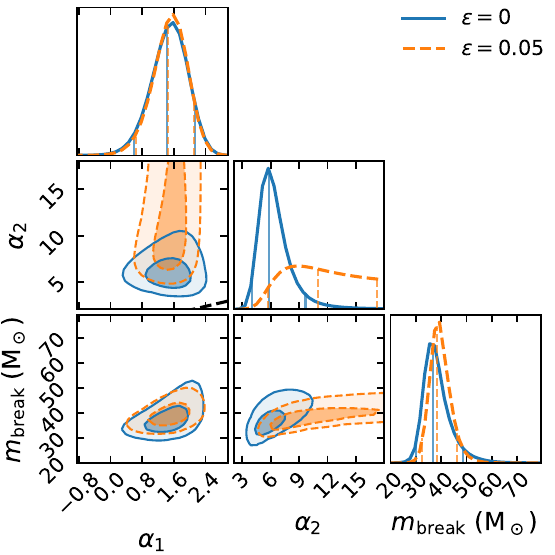}\label{fig:broken}}
    \subfloat[Power law + peak model]{
    \includegraphics[width=.333\linewidth]{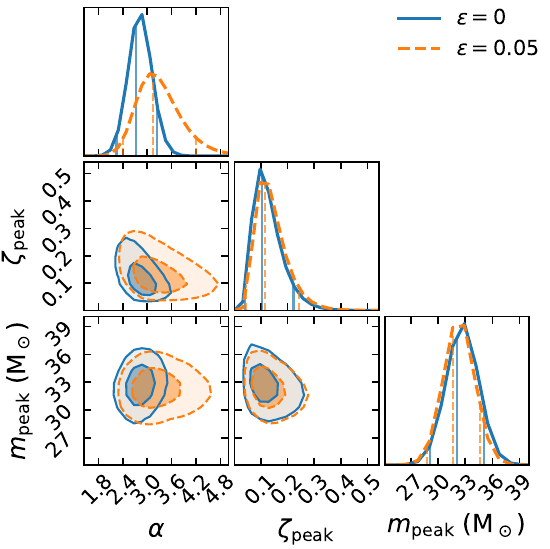}\label{fig:peak}}
    \caption{Adding a broad subpopulation $\lambda_0'$ with a fraction $\epsilon=0.05$ of the astrophysical rate affects the inferred parameters of the mass distribution. This is a major effect for the truncated power law model (a), moderate for the broken power law model (b) and minor for the power law + peak model (c). These constraints are derived using the GWTC-1 + IAS + Gold O3a sample of events; we find similar results with GWTC-1 + IAS + GWTC-2. \changed{Repeating the analysis with $\epsilon=0.1$ yields similar results as with $\epsilon=0.05$, corroborating that it is the freedom to accommodate outliers that drives these changes rather than the particular choice of $\epsilon$.}}
    \label{fig:mass_distribution}
\end{figure*}

The likelihood for this augmented model can be evaluated in post-processing from the same auxiliary quantities $w_i(\lambda', \lambda_0'), \overline{\mathcal{VT}}(\lambda'), \pastroi(\lambda_0)$ we use in the evaluation of Eq.~\eqref{eq:likelihood} (see \cite{Roulet2020}):
\begin{equation}
\begin{split}
    & P(\{d_i\} \mid \lambda, \lambda_0', \epsilon) \\
    &\propto \exp{-R[(1-\epsilon)\overline{\mathcal{VT}}(\lambda')
                 + \epsilon\overline{\mathcal{VT}}(\lambda_0')]} \\
    & \quad\times\prod_{i=1}^\ntrig \bigg\{\frac{R}{R_0}\big[w_i(\lambda', \lambda_0') (1 - \epsilon) + \epsilon\big] \pastroi(\lambda_0) \\
    &\quad\qquad\qquad+ 1 - \pastroi(\lambda_0)\bigg\} \, .
\end{split}
\end{equation}
Likewise, we can also extract the classification of each event as coming from the main component $\lambda$ or the broader component $\lambda'_0$: the probability that the $i$th event came from the $\lambda'_0$ population is
\begin{multline}\label{eq:p_outlier}
    p_{{\rm outlier}, i}(\lambda, \lambda'_0, \epsilon) \\
    = \frac{R\epsilon}{R[(1 - \epsilon)w_i(\lambda', \lambda_0')+\epsilon]+R_0(1/\pastroi(\lambda_0) - 1)} \, .
\end{multline}

We apply this procedure to three models of the mass distribution that are simplified versions of the \textsc{Truncated}, \textsc{Broken Power Law} and \textsc{Power Law + Peak} models studied in \citet{GWTC2_pop}. Our broken power law model is given by Eq.~\eqref{eq:broken_power_law}, with $\alpha_1, \alpha_2, \mbreak$ promoted to free parameters. Our truncated model corresponds to $\alpha_2 \to \infty$. Our power law + peak model corresponds to $\alpha_2 = \alpha_1$, plus the addition of a Gaussian component with mean $\mpeak$ and dispersion $\sigma = 5\,\msun$ that accounts for a fraction $\zetapeak$ of the total rate. In all three cases we assume a uniform distribution for $\chieff$, and identical distributions as in Section~\ref{ssec:spin_distribution} for the remaining parameters. With these choices, the models $\lambda'$ and $\lambda'_0$ only differ in the primary source-frame mass distribution, which will ease the interpretation of our results.

\begin{figure}[b!]
    \centering
    \includegraphics[width=\linewidth]{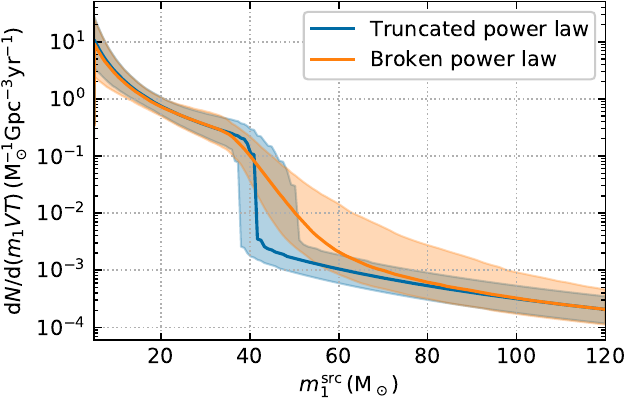}
    \caption{Mass distribution predicted by the models favored by the data: truncated power law or broken power law, in both cases with an additional broad subpopulation with a fraction $\epsilon=0.05$ of the total rate responsible for the shallow tail to high masses. These distributions feature a steepening around \SI{40}{\msun}, consistent with a step, and a flattening at higher mass. Note that we do not fit for the power-law index nor the normalization of the high-mass tail.}
    \label{fig:mass_predictive_distribution}
\end{figure}

Fig.~\ref{fig:mass_distribution} shows the constraints we obtain using the GWTC-1 + IAS + Gold O3a sample; these plots are largely unchanged if we use the GWTC-1 + IAS + GWTC-2 sample (not shown). For the case $\epsilon=0$ we find large quantitative agreement with \citet{GWTC2_pop} in the constraints for the corresponding model parameters; in particular, that the data favor a break with $\alpha_1>\alpha_2$ (region above the dashed line in Fig.~\ref{fig:broken}). When we set $\epsilon=0.05$, allowing these models to not fit all events, we find that the model parameter constraints are affected: for the truncated power-law model the effect is catastrophic, in the sense that the posteriors for $\epsilon=0$ and $0.05$ are inconsistent with each other; while for the broken power-law model there remains a region of overlap and for the power law + peak model the inferred parameters remain largely unaffected. This is in line with the discussion of \cite[figure 2]{GWTC2_pop} and suggests that the truncated power-law model with $\epsilon=0$ fails to describe the astrophysical distribution. It is interesting to note that, with $\epsilon=0.05$, the truncated and broken power law parametrizations are consistent with the same physical solution $\alpha \approx \alpha_1$, $\mmax \approx \mbreak$, $\alpha_2 \gg 1$, which exhibits a sharp step at $\mbreak$ and a tail that extends to high masses. The power law + peak parametrization cannot produce a step. We show these inferred distributions in Fig.~\ref{fig:mass_predictive_distribution} to further illustrate the point that both parametrizations give consistent answers, especially for the bulk of the distribution. Note that the differential merger rate is best constrained around $\mones \sim \SI{20}{\msun}$, where most observations lie \cite{Roulet2020}.

\begin{table}[t!]
    \centering
    \begin{tabular}{@{\extracolsep{4pt}}llll}
\toprule
                                     & $\epsilon$ & $\Delta \,\max \, \ln\, L$ &      $\Delta \,\ln \,Z$ \\
\midrule
\multirow{2}{*}{Truncated power law} &          0 &       $-7.4_{-0.3}^{+0.3}$ & $-6.21_{-0.19}^{+0.29}$ \\
                                     &       0.05 &                        $\phantom{-}0$ &          $\phantom{-}0$ \\
   \multirow{2}{*}{Broken power law} &          0 &    $-2.51_{-0.12}^{+0.14}$ & $-3.12_{-0.14}^{+0.13}$ \\
                                     &       0.05 &    $-0.23_{-0.11}^{+0.07}$ &  $\phantom{-}0.02_{-0.04}^{+0.03}$ \\
   \multirow{2}{*}{Power law + peak} &          0 &    $-1.03_{-0.17}^{+0.18}$ & $-3.18_{-0.13}^{+0.22}$ \\
                                     &       0.05 &     $\phantom{-}0.04_{-0.17}^{+0.21}$ & $-1.56_{-0.17}^{+0.15}$ \\
\bottomrule
\end{tabular}
    \caption{Scores for models of the primary mass distribution. Maximum log likelihood and log evidence for truncated power law and broken power law models, plus a fraction $\epsilon=0$ or $0.05$ of the population coming from a broad distribution $\lambda_0'$ per Eq.~\eqref{eq:wildcard}. The scores are referred to the preferred truncated power law model with $\epsilon=0.05$. In all three cases $\epsilon=0$ is disfavored, implying that the models struggle to accommodate all observations.}
    \label{tab:mass_models_scores}
\end{table}

In Table~\ref{tab:mass_models_scores} we report the maximum likelihood and evidence for each of the models studied. We find that, although the broken power law model outperforms the truncated model when $\epsilon=0$, both perform poorly relative to their $\epsilon = 0.05$ counterparts. This suggests that neither is a good description of the mass distribution.
The power law + peak model achieves similar scores as the broken power law model with $\epsilon=0$, but it gets only a slight improvement from $\epsilon=0.05$, thus getting similarly disfavored. Among all the variations, thus, the preferred models in terms of evidence are either the truncated or broken power law with $\epsilon=0.05$, i.e.\ with a small fraction of events in a broad tail that extends to high masses. The fact that these two models achieve similar likelihood and evidence, together with the above observation that they are consistent with the same physical solution, suggests that both are comparably good descriptions of the bulk of the distribution and their different scores for $\epsilon=0$ are driven by the few outlier events. This is confirmed in Fig.~\ref{fig:mass_predictive_distribution}. Comparing the $\epsilon=0$ entries in Table~\ref{tab:scores} to \citet[table 2]{GWTC2_pop}, we find agreement in that the truncated power law model is rejected, however, \citet{GWTC2_pop} find a preference for the power law + peak model over the broken power law, which we instead find comparable. Some differences are expected because, for simplicity, in our implementation of these models we fixed or omitted some parameters, so the models and associated phase spaces are not equivalent.

We can get some insight by inspecting the probabilities $p_{\rm outlier}$ of coming from the broad subpopulation $\lambda_0'$ assigned to each event, which we report in Table~\ref{tab:p_outlier}. Events with a high value of $p_{\rm outlier}$ are better explained by the broad subpopulation and drive a preference for $\epsilon \neq 0$. However, note that even if the true astrophysical population was well described by the parametrization $\lambda$, in a catalog of many events some are bound to be in the tail of the distribution and might individually be better described by a broad distribution. The expected distribution of $p_{\rm outlier}$ under a model $\lambda$ is hard to compute, which is why we do not use the values of $p_{\rm outlier}$ as a quantitative model test. This said, it is apparent that GW190521 is an extreme outlier of the truncated power law model, and there are two other events that are in some tension. For the broken power law model, GW190521 is in some tension but the other values of $p_{\rm outlier}$ are milder. For the power law + peak model, no single event is a strong outlier. 

\begin{table}
    \centering
    \begin{tabular}{l
                >{\centering\arraybackslash} p{1.6cm}
                >{\centering\arraybackslash}p{1.6cm}
                >{\centering\arraybackslash}p{1.6cm}}
        \toprule
        {} &  Truncated\newline power law &  Broken\newline power law &  Power law\newline + peak \\
        \midrule
        GW190521         &    1.00 & 0.94 & 0.68 \\
        GW190602\_175927 &    0.95 & 0.72 & 0.66 \\
        GW190706\_222641 &    0.88 & 0.72 & 0.75 \\
        GW190519\_153544 &    0.76 & 0.54 & 0.59 \\
        GW190929\_012149 &    0.57 & 0.46 & 0.51 \\
        GW190620\_030421 &    0.43 & 0.34 & 0.47 \\
        GW190701\_203306 &    0.33 & 0.19 & 0.29 \\
        GW190413\_134308 &    0.27 & 0.25 & 0.31 \\
        \bottomrule
    \end{tabular}
    \caption{Probability that each event is a model outlier, as defined in Eq.~\eqref{eq:p_outlier} and marginalized over model parameters $\lambda$, with $\epsilon=0.05$, for the mass models studied. Only events in the gold sample with the highest values of $p_{\rm outlier}$ are shown, for brevity. Note that this naturally selects the events with highest primary mass.}
    \label{tab:p_outlier}
\end{table}

Another interesting effect is that GW170817A, a candidate event with $\mones = 56^{+16}_{-10}\,\msun$ and a rather low false-alarm rate of $1/(\SI{36}{O2})$ observing runs \cite{fishing}, had an estimated probability of astrophysical origin marginalized over population parameters of $\overline\pastro = 0.07$, under the truncated power law model favored after O1 and O2 \cite{Roulet2020}. This low value was driven by the lack of observations of other events with similar properties, mainly mass. Under the newly favored models, it has a moderately different $\overline\pastro = 0.22$ for the truncated power law and 0.26 for the broken power law, both with $\epsilon=0.05$. This showcases that $\pastro$ values for marginal events in the tails of the distribution are bound to get updated as our knowledge of the population improves.

To summarize, Table~\ref{tab:mass_models_scores}, Figs.~\ref{fig:mass_distribution} and \ref{fig:mass_predictive_distribution} suggest that the mass distribution exhibits a steepening around $40\,\msun$ and an extended, shallow high-mass tail. From Table~\ref{tab:p_outlier} we conclude that the need for this tail is dominated by GW190521, so at this point we do not attempt to model its shape based on a single event. Future data releases will allow to probe these features in the mass distribution.

\section{Conclusions}
\label{sec:conclusions}

We have investigated the properties of the effective spin and primary mass distributions of binary black holes identified in the GWTC-1~\cite{GWTC-1}, GWTC-2~\cite{GWTC2}, and IAS O1--O2~\cite{Venumadhav:2019tad, Zackay:2019tzo, Venumadhav:2019lyq, fishing} event catalogs. Our study involved re-analyzing all binary black hole signals with the recently developed \xphm\ waveform model~\cite{Pratten2020}, which includes orbital precession and higher-order modes.

We designed a parametric model of the $\chieff$ distribution which has three components --- with negative, approximately zero and positive $\chieff$ --- to test some general predictions of the dynamic and isolated formation channels for merging binary black holes. Namely, dynamical formation channels predict a $\chieff$ distribution that is symmetric about $0$, while negative $\chieff$ (i.e., large spin--orbit misalignment) should be very rare for isolated field binaries.
Interestingly, we found that a symmetric distribution is disfavored: the data suggests that the number of positive $\chieff$ events is larger than that with negative $\chieff$ at 95\% credibility. Although the evidence at this point is not conclusive, this simple test is already becoming powerful enough to hint that not all binary black holes are dynamically assembled, in agreement with other analyses of these data \cite{GWTC2_pop, Zevin2021, Bouffanais2021}. The number of detections is expected to roughly double with the forthcoming release of the O3b catalog, which should settle this question if the same trend continues.

Moreover, we find no evidence for negative $\chieff$ in the population. This result is in tension with Ref.~\cite{GWTC2_pop}; we attribute the discrepancy to the different parametrizations of the spin distribution chosen. We were able to reproduce the results of Ref.~\cite{GWTC2_pop} with a Gaussian model for $\chieff$, but found that this model fares worse at describing the features in the spin distribution, in particular, a large concentration of events near $\chieff=0$. Our conclusion is in agreement with a model-free inspection of the empirical $\chieff$ distribution, which suggests that all events with significant support at $\chieff<0$ are consistent with coming from a population with $\chieff=0$. Therefore, we conclude that the observed effective spin distribution does not rule out that all observations are explained by isolated binary formation.

Regarding the distribution of primary masses, we confirmed the result of \cite{GWTC2_pop} that a truncated power law fails to describe the observations. Moreover, we found evidence that a broken power law model or a power law plus a Gaussian peak, which assume a continuous distribution, compare poorly to a model in which a small fraction of the events comes from a broad subpopulation, with an extended tail at high masses. This suggests that the tail of the mass distribution has interesting features that will be probed with the coming data releases.

\section*{Acknowledgements}

We are grateful to Ilya Mandel, Will Farr, Davide Gerosa and the attendees of the gravitational wave meeting at the Center for Computational Astrophysics for helpful comments and discussion. 

HSC gratefully acknowledges support from the Rubicon Fellowship awarded by the Netherlands Organisation for Scientific Research (NWO). SO acknowledges support from the National Science Foundation Graduate Research Fellowship Program under Grant No. DGE-2039656. Any opinions, findings, and conclusions or recommendations expressed in this material are those of the authors and do not necessarily reflect the views of the National Science Foundation. LD acknowledges support from the Michael M. Garland startup research grant at the University of California, Berkeley. TV acknowledges support by the National Science Foundation under Grant No. 2012086. BZ is supported by a research grant from the Ruth and Herman Albert Scholarship Program for New Scientists. MZ is supported by NSF grants PHY-1820775 the Canadian Institute for Advanced Research (CIFAR) Program on Gravity and the Extreme Universe and the Simons Foundation Modern Inflationary Cosmology initiative. 

This research has made use of data, software and/or web tools obtained from the Gravitational Wave Open Science Center (\url{https://www.gw-openscience.org/}), a service of LIGO Laboratory, the LIGO Scientific Collaboration and the Virgo Collaboration. LIGO Laboratory and Advanced LIGO are funded by the United States National Science Foundation (NSF) as well as the Science and Technology Facilities Council (STFC) of the United Kingdom, the Max-Planck-Society (MPS), and the State of Niedersachsen/Germany for support of the construction of Advanced LIGO and construction and operation of the GEO600 detector. Additional support for Advanced LIGO was provided by the Australian Research Council. Virgo is funded, through the European Gravitational Observatory (EGO), by the French Centre National de Recherche Scientifique (CNRS), the Italian Istituto Nazionale di Fisica Nucleare (INFN) and the Dutch Nikhef, with contributions by institutions from Belgium, Germany, Greece, Hungary, Ireland, Japan, Monaco, Poland, Portugal, Spain.

\appendix 

\section{Sample Selection} 
\label{app:gold_sample}

In this appendix, we inventorize the binary black hole mergers used in this work, which are listed in Table~\ref{tab:events}. We define the gold sample (third column of Table~\ref{tab:events}) as the set of events that (i) were detected by at least two search pipelines with a $\rm FAR < \SI{0.1}{yr^{-1}}$ (fourth column); and (ii) on strain data that are free of non-Gaussian transient noise (fifth column). We consider the following pipelines: cWB \cite{GWTC2}, GstLAL \cite{GWTC2}, PyCBC \cite{GWTC2, Nitz:2018imz, Nitz:2019hdf}, PyCBC BBH \cite{GWTC2, Nitz:2019hdf}, and IAS \cite{Venumadhav:2019lyq}. 
Events with non-Gaussian artifacts are reported in \cite[table V]{GWTC2_pop}. We do not include GW190814 in the GWTC-1 + IAS + Gold O3a sample because it was detected near non-Gaussian transient noise \cite{GWTC2}. Nor do we include GW190814 in the GWTC-1 + IAS + GWTC-2 sample (Section~\ref{ssec:spin_distribution}) as it was not included in the main GWTC-2 population analysis due to being an outlier in the mass ratio distribution \cite{GWTC2_pop}. For events in the O1 and O2 observing runs, $\pastro(\lambda_0)$ is computed in \cite{Roulet2020}. For events in O3a, it is taken at face value from \cite{GWTC2} as the maximum $\pastro$ over pipelines, and may not accurately correspond to the model $\lambda_0$.

While the present work was being completed, \citet{3OGC} reported their analysis of the O3a data, providing independent confirmation of all the sources reported in GWTC-2 except for GW190426\_152155 and GW190909\_114149, and further finding four previously unreported events. We defer the inclusion of these results to future work. Including this catalog, the two-pipeline condition would be fulfilled by most of the O3a events in Table~\ref{tab:events}, thereby enlarging the gold sample. Still, note that the sample restriction did not change the qualitative conclusions of our analysis. 

\begin{table}
    \centering
    \begin{tabular}{llcccr}
    \toprule
    Run &            Name &  Gold &  $\geq 2$ pip. &  Clean &  $p_{\rm astro}(\lambda_0)$ \\
    \midrule
     O1 &        GW150914 &  \checkmark & \checkmark &   \checkmark & 1.00 \\
        &        GW151012 &  \checkmark & \checkmark &   \checkmark & 1.00 \\
        &        GW151226 &  \checkmark & \checkmark &   \checkmark & 1.00 \\
        &        GW151216 &             &            &   \checkmark & 0.50 \\
     O2 &        GW170823 &  \checkmark & \checkmark &   \checkmark & 1.00 \\
        &        GW170809 &  \checkmark & \checkmark &   \checkmark & 1.00 \\
        &        GW170729 &  \checkmark & \checkmark &   \checkmark & 1.00 \\
        &        GW170814 &  \checkmark & \checkmark &   \checkmark & 1.00 \\
        &        GW170104 &  \checkmark & \checkmark &   \checkmark & 1.00 \\
        &        GW170727 &  \checkmark & \checkmark &   \checkmark & 0.99 \\
        &        GW170121 &  \checkmark & \checkmark &   \checkmark & 0.97 \\
        &        GW170304 &             &            &   \checkmark & 1.00 \\
        &        GW170818 &  \checkmark & \checkmark &   \checkmark & 0.92 \\
        &         170412B &             &            &   \checkmark & 0.02 \\
        &        GW170403 &             &            &   \checkmark & 0.61 \\
        &        GW170425 &             &            &   \checkmark & 0.60 \\
        &        GW170202 &             &            &   \checkmark & 0.61 \\
        &       GW170817A &             &            &   \checkmark & 0.74 \\
        &        GW170608 &  \checkmark & \checkmark &   \checkmark & 1.00 \\
    O3a & GW190408\_181802 &  \checkmark & \checkmark &   \checkmark & 1.00 \\
        &        GW190412 &  \checkmark & \checkmark &   \checkmark & 1.00 \\
        & GW190413\_052954 &             &            &   \checkmark & 0.98 \\
        & GW190413\_134308 &             &            &              & 0.98 \\
        & GW190421\_213856 &  \checkmark & \checkmark &   \checkmark & 1.00 \\
        & GW190424\_180648 &             &            &              & 0.91 \\
        & GW190503\_185404 &             & \checkmark &              & 1.00 \\
        & GW190512\_180714 &  \checkmark & \checkmark &   \checkmark & 1.00 \\
        & GW190513\_205428 &             & \checkmark &              & 1.00 \\
        & GW190514\_065416 &             &            &              & 0.96 \\
        & GW190517\_055101 &  \checkmark & \checkmark &   \checkmark & 1.00 \\
        & GW190519\_153544 &  \checkmark & \checkmark &   \checkmark & 1.00 \\
        &        GW190521 &  \checkmark & \checkmark &   \checkmark & 1.00 \\
        & GW190521\_074359 &  \checkmark & \checkmark &   \checkmark & 1.00 \\
        & GW190527\_092055 &             &            &   \checkmark & 0.99 \\
        & GW190602\_175927 &  \checkmark & \checkmark &   \checkmark & 1.00 \\
        & GW190620\_030421 &             &            &   \checkmark & 1.00 \\
        & GW190630\_185205 &             &            &   \checkmark & 1.00 \\
        & GW190701\_203306 &             &            &              & 1.00 \\
        & GW190706\_222641 &  \checkmark & \checkmark &   \checkmark & 1.00 \\
        & GW190707\_093326 &  \checkmark & \checkmark &   \checkmark & 1.00 \\
        & GW190708\_232457 &             &            &   \checkmark & 1.00 \\
        & GW190719\_215514 &             &            &   \checkmark & 0.82 \\
        & GW190720\_000836 &  \checkmark & \checkmark &   \checkmark & 1.00 \\
        & GW190727\_060333 &             & \checkmark &              & 1.00 \\
        & GW190728\_064510 &  \checkmark & \checkmark &   \checkmark & 1.00 \\
        & GW190731\_140936 &             &            &   \checkmark & 0.97 \\
        & GW190803\_022701 &  \checkmark & \checkmark &   \checkmark & 0.99 \\
        & GW190828\_063405 &  \checkmark & \checkmark &   \checkmark & 1.00 \\
        & GW190828\_065509 &  \checkmark & \checkmark &   \checkmark & 1.00 \\
        & GW190909\_114149 &             &            &   \checkmark & 0.89 \\
        & GW190910\_112807 &             &            &   \checkmark & 1.00 \\
        & GW190915\_235702 &  \checkmark & \checkmark &   \checkmark & 1.00 \\
        & GW190924\_021846 &             & \checkmark &              & 1.00 \\
        & GW190929\_012149 &             &            &   \checkmark & 1.00 \\
        & GW190930\_133541 &  \checkmark & \checkmark &   \checkmark & 1.00 \\
    \bottomrule
    \end{tabular}
    \caption{Binary black hole events used in this work. Checkmarks from the third to fifth columns indicate events that are in the gold sample, were identified by at least two pipelines with $\text{IFAR} > \SI{10}{yr}$, and were observed in the absence of glitches, respectively. The $p_{\rm astro}$ values shown here are evaluated with the reference model $\lambda_0$ described in Eq.~(\ref{eq:f0}).}
    \label{tab:events}
\end{table}

\clearpage

\bibliographystyle{apsrev4-1}
\bibliography{O1O3aPop}

\begin{thebibliography}{106}%
\makeatletter
\providecommand \@ifxundefined [1]{%
 \@ifx{#1\undefined}
}%
\providecommand \@ifnum [1]{%
 \ifnum #1\expandafter \@firstoftwo
 \else \expandafter \@secondoftwo
 \fi
}%
\providecommand \@ifx [1]{%
 \ifx #1\expandafter \@firstoftwo
 \else \expandafter \@secondoftwo
 \fi
}%
\providecommand \natexlab [1]{#1}%
\providecommand \enquote  [1]{``#1''}%
\providecommand \bibnamefont  [1]{#1}%
\providecommand \bibfnamefont [1]{#1}%
\providecommand \citenamefont [1]{#1}%
\providecommand \href@noop [0]{\@secondoftwo}%
\providecommand \href [0]{\begingroup \@sanitize@url \@href}%
\providecommand \@href[1]{\@@startlink{#1}\@@href}%
\providecommand \@@href[1]{\endgroup#1\@@endlink}%
\providecommand \@sanitize@url [0]{\catcode `\\12\catcode `\$12\catcode
  `\&12\catcode `\#12\catcode `\^12\catcode `\_12\catcode `\%12\relax}%
\providecommand \@@startlink[1]{}%
\providecommand \@@endlink[0]{}%
\providecommand \url  [0]{\begingroup\@sanitize@url \@url }%
\providecommand \@url [1]{\endgroup\@href {#1}{\urlprefix }}%
\providecommand \urlprefix  [0]{URL }%
\providecommand \Eprint [0]{\href }%
\providecommand \doibase [0]{http://dx.doi.org/}%
\providecommand \selectlanguage [0]{\@gobble}%
\providecommand \bibinfo  [0]{\@secondoftwo}%
\providecommand \bibfield  [0]{\@secondoftwo}%
\providecommand \translation [1]{[#1]}%
\providecommand \BibitemOpen [0]{}%
\providecommand \bibitemStop [0]{}%
\providecommand \bibitemNoStop [0]{.\EOS\space}%
\providecommand \EOS [0]{\spacefactor3000\relax}%
\providecommand \BibitemShut  [1]{\csname bibitem#1\endcsname}%
\let\auto@bib@innerbib\@empty
\bibitem [{\citenamefont {Abbott}\ \emph
  {et~al.}(2020{\natexlab{a}})\citenamefont {Abbott} \emph {et~al.}}]{GWTC2}%
  \BibitemOpen
  \bibfield  {author} {\bibinfo {author} {\bibfnamefont {R.}~\bibnamefont
  {Abbott}} \emph {et~al.},\ }\href@noop {} {\enquote {\bibinfo {title}
  {{GWTC}-2: Compact binary coalescences observed by {LIGO} and {V}irgo during
  the first half of the third observing run},}\ } (\bibinfo {year}
  {2020}{\natexlab{a}}),\ \Eprint {http://arxiv.org/abs/arXiv:2010.14527}
  {arXiv:2010.14527} \BibitemShut {NoStop}%
\bibitem [{\citenamefont {Abbott}\ \emph
  {et~al.}(2019{\natexlab{a}})\citenamefont {Abbott} \emph {et~al.}}]{GWTC-1}%
  \BibitemOpen
  \bibfield  {author} {\bibinfo {author} {\bibfnamefont {B.~P.}\ \bibnamefont
  {Abbott}} \emph {et~al.} (\bibinfo {collaboration} {LIGO Scientific
  Collaboration and Virgo Collaboration}),\ }\href {\doibase
  10.1103/PhysRevX.9.031040} {\bibfield  {journal} {\bibinfo  {journal} {Phys.
  Rev. X}\ }\textbf {\bibinfo {volume} {9}},\ \bibinfo {pages} {031040}
  (\bibinfo {year} {2019}{\natexlab{a}})}\BibitemShut {NoStop}%
\bibitem [{\citenamefont {Nitz}\ \emph {et~al.}(2019)\citenamefont {Nitz},
  \citenamefont {Capano}, \citenamefont {Nielsen}, \citenamefont {Reyes},
  \citenamefont {White}, \citenamefont {Brown},\ and\ \citenamefont
  {Krishnan}}]{Nitz:2018imz}%
  \BibitemOpen
  \bibfield  {author} {\bibinfo {author} {\bibfnamefont {A.~H.}\ \bibnamefont
  {Nitz}}, \bibinfo {author} {\bibfnamefont {C.}~\bibnamefont {Capano}},
  \bibinfo {author} {\bibfnamefont {A.~B.}\ \bibnamefont {Nielsen}}, \bibinfo
  {author} {\bibfnamefont {S.}~\bibnamefont {Reyes}}, \bibinfo {author}
  {\bibfnamefont {R.}~\bibnamefont {White}}, \bibinfo {author} {\bibfnamefont
  {D.~A.}\ \bibnamefont {Brown}}, \ and\ \bibinfo {author} {\bibfnamefont
  {B.}~\bibnamefont {Krishnan}},\ }\href {\doibase 10.3847/1538-4357/ab0108}
  {\bibfield  {journal} {\bibinfo  {journal} {Astrophys. J.}\ }\textbf
  {\bibinfo {volume} {872}},\ \bibinfo {pages} {195} (\bibinfo {year}
  {2019})},\ \Eprint {http://arxiv.org/abs/1811.01921} {arXiv:1811.01921
  [gr-qc]} \BibitemShut {NoStop}%
\bibitem [{\citenamefont {Venumadhav}\ \emph
  {et~al.}(2019{\natexlab{a}})\citenamefont {Venumadhav}, \citenamefont
  {Zackay}, \citenamefont {Roulet}, \citenamefont {Dai},\ and\ \citenamefont
  {Zaldarriaga}}]{Venumadhav:2019tad}%
  \BibitemOpen
  \bibfield  {author} {\bibinfo {author} {\bibfnamefont {T.}~\bibnamefont
  {Venumadhav}}, \bibinfo {author} {\bibfnamefont {B.}~\bibnamefont {Zackay}},
  \bibinfo {author} {\bibfnamefont {J.}~\bibnamefont {Roulet}}, \bibinfo
  {author} {\bibfnamefont {L.}~\bibnamefont {Dai}}, \ and\ \bibinfo {author}
  {\bibfnamefont {M.}~\bibnamefont {Zaldarriaga}},\ }\href {\doibase
  10.1103/PhysRevD.100.023011} {\bibfield  {journal} {\bibinfo  {journal}
  {Phys. Rev. D}\ }\textbf {\bibinfo {volume} {100}},\ \bibinfo {pages}
  {023011} (\bibinfo {year} {2019}{\natexlab{a}})},\ \Eprint
  {http://arxiv.org/abs/1902.10341} {arXiv:1902.10341 [astro-ph.IM]}
  \BibitemShut {NoStop}%
\bibitem [{\citenamefont {Zackay}\ \emph
  {et~al.}(2019{\natexlab{a}})\citenamefont {Zackay}, \citenamefont
  {Venumadhav}, \citenamefont {Dai}, \citenamefont {Roulet},\ and\
  \citenamefont {Zaldarriaga}}]{Zackay:2019tzo}%
  \BibitemOpen
  \bibfield  {author} {\bibinfo {author} {\bibfnamefont {B.}~\bibnamefont
  {Zackay}}, \bibinfo {author} {\bibfnamefont {T.}~\bibnamefont {Venumadhav}},
  \bibinfo {author} {\bibfnamefont {L.}~\bibnamefont {Dai}}, \bibinfo {author}
  {\bibfnamefont {J.}~\bibnamefont {Roulet}}, \ and\ \bibinfo {author}
  {\bibfnamefont {M.}~\bibnamefont {Zaldarriaga}},\ }\href {\doibase
  10.1103/PhysRevD.100.023007} {\bibfield  {journal} {\bibinfo  {journal}
  {Phys. Rev. D}\ }\textbf {\bibinfo {volume} {100}},\ \bibinfo {pages}
  {023007} (\bibinfo {year} {2019}{\natexlab{a}})},\ \Eprint
  {http://arxiv.org/abs/1902.10331} {arXiv:1902.10331 [astro-ph.HE]}
  \BibitemShut {NoStop}%
\bibitem [{\citenamefont {Venumadhav}\ \emph
  {et~al.}(2020{\natexlab{a}})\citenamefont {Venumadhav}, \citenamefont
  {Zackay}, \citenamefont {Roulet}, \citenamefont {Dai},\ and\ \citenamefont
  {Zaldarriaga}}]{Venumadhav:2019lyq}%
  \BibitemOpen
  \bibfield  {author} {\bibinfo {author} {\bibfnamefont {T.}~\bibnamefont
  {Venumadhav}}, \bibinfo {author} {\bibfnamefont {B.}~\bibnamefont {Zackay}},
  \bibinfo {author} {\bibfnamefont {J.}~\bibnamefont {Roulet}}, \bibinfo
  {author} {\bibfnamefont {L.}~\bibnamefont {Dai}}, \ and\ \bibinfo {author}
  {\bibfnamefont {M.}~\bibnamefont {Zaldarriaga}},\ }\href {\doibase
  10.1103/PhysRevD.101.083030} {\bibfield  {journal} {\bibinfo  {journal}
  {Phys. Rev. D}\ }\textbf {\bibinfo {volume} {101}},\ \bibinfo {pages}
  {083030} (\bibinfo {year} {2020}{\natexlab{a}})},\ \Eprint
  {http://arxiv.org/abs/1904.07214} {arXiv:1904.07214 [astro-ph.HE]}
  \BibitemShut {NoStop}%
\bibitem [{\citenamefont {Zackay}\ \emph
  {et~al.}(2019{\natexlab{b}})\citenamefont {Zackay}, \citenamefont {Dai},
  \citenamefont {Venumadhav}, \citenamefont {Roulet},\ and\ \citenamefont
  {Zaldarriaga}}]{fishing}%
  \BibitemOpen
  \bibfield  {author} {\bibinfo {author} {\bibfnamefont {B.}~\bibnamefont
  {Zackay}}, \bibinfo {author} {\bibfnamefont {L.}~\bibnamefont {Dai}},
  \bibinfo {author} {\bibfnamefont {T.}~\bibnamefont {Venumadhav}}, \bibinfo
  {author} {\bibfnamefont {J.}~\bibnamefont {Roulet}}, \ and\ \bibinfo {author}
  {\bibfnamefont {M.}~\bibnamefont {Zaldarriaga}},\ }\href@noop {} {\enquote
  {\bibinfo {title} {{Detecting Gravitational Waves With Disparate Detector
  Responses: Two New Binary Black Hole Mergers}},}\ } (\bibinfo {year}
  {2019}{\natexlab{b}}),\ \Eprint {http://arxiv.org/abs/1910.09528}
  {arXiv:1910.09528 [astro-ph.HE]} \BibitemShut {NoStop}%
\bibitem [{\citenamefont {Nitz}\ \emph {et~al.}(2020)\citenamefont {Nitz},
  \citenamefont {Dent}, \citenamefont {Davies}, \citenamefont {Kumar},
  \citenamefont {Capano}, \citenamefont {Harry}, \citenamefont {Mozzon},
  \citenamefont {Nuttall}, \citenamefont {Lundgren},\ and\ \citenamefont
  {T\'apai}}]{Nitz:2019hdf}%
  \BibitemOpen
  \bibfield  {author} {\bibinfo {author} {\bibfnamefont {A.~H.}\ \bibnamefont
  {Nitz}}, \bibinfo {author} {\bibfnamefont {T.}~\bibnamefont {Dent}}, \bibinfo
  {author} {\bibfnamefont {G.~S.}\ \bibnamefont {Davies}}, \bibinfo {author}
  {\bibfnamefont {S.}~\bibnamefont {Kumar}}, \bibinfo {author} {\bibfnamefont
  {C.~D.}\ \bibnamefont {Capano}}, \bibinfo {author} {\bibfnamefont
  {I.}~\bibnamefont {Harry}}, \bibinfo {author} {\bibfnamefont
  {S.}~\bibnamefont {Mozzon}}, \bibinfo {author} {\bibfnamefont
  {L.}~\bibnamefont {Nuttall}}, \bibinfo {author} {\bibfnamefont
  {A.}~\bibnamefont {Lundgren}}, \ and\ \bibinfo {author} {\bibfnamefont
  {M.}~\bibnamefont {T\'apai}},\ }\href {\doibase 10.3847/1538-4357/ab733f}
  {\bibfield  {journal} {\bibinfo  {journal} {Astrophys. J.}\ }\textbf
  {\bibinfo {volume} {891}},\ \bibinfo {pages} {123} (\bibinfo {year}
  {2020})},\ \Eprint {http://arxiv.org/abs/1910.05331} {arXiv:1910.05331
  [astro-ph.HE]} \BibitemShut {NoStop}%
\bibitem [{\citenamefont {Abbott}\ \emph
  {et~al.}(2020{\natexlab{b}})\citenamefont {Abbott} \emph
  {et~al.}}]{GWTC2_pop}%
  \BibitemOpen
  \bibfield  {author} {\bibinfo {author} {\bibfnamefont {R.}~\bibnamefont
  {Abbott}} \emph {et~al.},\ }\href@noop {} {\enquote {\bibinfo {title}
  {Population properties of compact objects from the second {LIGO}--{V}irgo
  {G}ravitational-{W}ave {T}ransient {C}atalog},}\ } (\bibinfo {year}
  {2020}{\natexlab{b}}),\ \Eprint {http://arxiv.org/abs/arXiv:2010.14533}
  {arXiv:2010.14533} \BibitemShut {NoStop}%
\bibitem [{\citenamefont {Wong}\ \emph {et~al.}(2021)\citenamefont {Wong},
  \citenamefont {Breivik}, \citenamefont {Kremer},\ and\ \citenamefont
  {Callister}}]{Wong2021}%
  \BibitemOpen
  \bibfield  {author} {\bibinfo {author} {\bibfnamefont {K.~W.~K.}\
  \bibnamefont {Wong}}, \bibinfo {author} {\bibfnamefont {K.}~\bibnamefont
  {Breivik}}, \bibinfo {author} {\bibfnamefont {K.}~\bibnamefont {Kremer}}, \
  and\ \bibinfo {author} {\bibfnamefont {T.}~\bibnamefont {Callister}},\ }\href
  {\doibase 10.1103/PhysRevD.103.083021} {\bibfield  {journal} {\bibinfo
  {journal} {Phys. Rev. D}\ }\textbf {\bibinfo {volume} {103}},\ \bibinfo
  {pages} {083021} (\bibinfo {year} {2021})}\BibitemShut {NoStop}%
\bibitem [{\citenamefont {Zevin}\ \emph {et~al.}(2021)\citenamefont {Zevin},
  \citenamefont {Bavera}, \citenamefont {Berry}, \citenamefont {Kalogera},
  \citenamefont {Fragos}, \citenamefont {Marchant}, \citenamefont {Rodriguez},
  \citenamefont {Antonini}, \citenamefont {Holz},\ and\ \citenamefont
  {Pankow}}]{Zevin2021}%
  \BibitemOpen
  \bibfield  {author} {\bibinfo {author} {\bibfnamefont {M.}~\bibnamefont
  {Zevin}}, \bibinfo {author} {\bibfnamefont {S.~S.}\ \bibnamefont {Bavera}},
  \bibinfo {author} {\bibfnamefont {C.~P.~L.}\ \bibnamefont {Berry}}, \bibinfo
  {author} {\bibfnamefont {V.}~\bibnamefont {Kalogera}}, \bibinfo {author}
  {\bibfnamefont {T.}~\bibnamefont {Fragos}}, \bibinfo {author} {\bibfnamefont
  {P.}~\bibnamefont {Marchant}}, \bibinfo {author} {\bibfnamefont {C.~L.}\
  \bibnamefont {Rodriguez}}, \bibinfo {author} {\bibfnamefont {F.}~\bibnamefont
  {Antonini}}, \bibinfo {author} {\bibfnamefont {D.~E.}\ \bibnamefont {Holz}},
  \ and\ \bibinfo {author} {\bibfnamefont {C.}~\bibnamefont {Pankow}},\ }\href
  {\doibase 10.3847/1538-4357/abe40e} {\bibfield  {journal} {\bibinfo
  {journal} {The Astrophysical Journal}\ }\textbf {\bibinfo {volume} {910}},\
  \bibinfo {pages} {152} (\bibinfo {year} {2021})}\BibitemShut {NoStop}%
\bibitem [{\citenamefont {Bouffanais}\ \emph {et~al.}(2021)\citenamefont
  {Bouffanais}, \citenamefont {Mapelli}, \citenamefont {Santoliquido},
  \citenamefont {Giacobbo}, \citenamefont {Carlo}, \citenamefont {Rastello},
  \citenamefont {Artale},\ and\ \citenamefont {Iorio}}]{Bouffanais2021}%
  \BibitemOpen
  \bibfield  {author} {\bibinfo {author} {\bibfnamefont {Y.}~\bibnamefont
  {Bouffanais}}, \bibinfo {author} {\bibfnamefont {M.}~\bibnamefont {Mapelli}},
  \bibinfo {author} {\bibfnamefont {F.}~\bibnamefont {Santoliquido}}, \bibinfo
  {author} {\bibfnamefont {N.}~\bibnamefont {Giacobbo}}, \bibinfo {author}
  {\bibfnamefont {U.~N.~D.}\ \bibnamefont {Carlo}}, \bibinfo {author}
  {\bibfnamefont {S.}~\bibnamefont {Rastello}}, \bibinfo {author}
  {\bibfnamefont {M.~C.}\ \bibnamefont {Artale}}, \ and\ \bibinfo {author}
  {\bibfnamefont {G.}~\bibnamefont {Iorio}},\ }\href@noop {} {\enquote
  {\bibinfo {title} {New insights on binary black hole formation channels after
  {GWTC}-2: young star clusters versus isolated binaries},}\ } (\bibinfo {year}
  {2021}),\ \Eprint {http://arxiv.org/abs/arXiv:2102.12495} {arXiv:2102.12495}
  \BibitemShut {NoStop}%
\bibitem [{\citenamefont {Zwart}\ and\ \citenamefont
  {McMillan}(1999)}]{zwart1999black}%
  \BibitemOpen
  \bibfield  {author} {\bibinfo {author} {\bibfnamefont {S.~F.~P.}\
  \bibnamefont {Zwart}}\ and\ \bibinfo {author} {\bibfnamefont {S.~L.}\
  \bibnamefont {McMillan}},\ }\href@noop {} {\bibfield  {journal} {\bibinfo
  {journal} {The Astrophysical Journal Letters}\ }\textbf {\bibinfo {volume}
  {528}},\ \bibinfo {pages} {L17} (\bibinfo {year} {1999})}\BibitemShut
  {NoStop}%
\bibitem [{\citenamefont {O’Leary}\ \emph {et~al.}(2006)\citenamefont
  {O’Leary}, \citenamefont {Rasio}, \citenamefont {Fregeau}, \citenamefont
  {Ivanova},\ and\ \citenamefont {O'Shaughnessy}}]{o2006binary}%
  \BibitemOpen
  \bibfield  {author} {\bibinfo {author} {\bibfnamefont {R.~M.}\ \bibnamefont
  {O’Leary}}, \bibinfo {author} {\bibfnamefont {F.~A.}\ \bibnamefont
  {Rasio}}, \bibinfo {author} {\bibfnamefont {J.~M.}\ \bibnamefont {Fregeau}},
  \bibinfo {author} {\bibfnamefont {N.}~\bibnamefont {Ivanova}}, \ and\
  \bibinfo {author} {\bibfnamefont {R.}~\bibnamefont {O'Shaughnessy}},\
  }\href@noop {} {\bibfield  {journal} {\bibinfo  {journal} {The Astrophysical
  Journal}\ }\textbf {\bibinfo {volume} {637}},\ \bibinfo {pages} {937}
  (\bibinfo {year} {2006})}\BibitemShut {NoStop}%
\bibitem [{\citenamefont {Sadowski}\ \emph {et~al.}(2008)\citenamefont
  {Sadowski}, \citenamefont {Belczynski}, \citenamefont {Bulik}, \citenamefont
  {Ivanova}, \citenamefont {Rasio},\ and\ \citenamefont
  {O'Shaughnessy}}]{sadowski2008total}%
  \BibitemOpen
  \bibfield  {author} {\bibinfo {author} {\bibfnamefont {A.}~\bibnamefont
  {Sadowski}}, \bibinfo {author} {\bibfnamefont {K.}~\bibnamefont
  {Belczynski}}, \bibinfo {author} {\bibfnamefont {T.}~\bibnamefont {Bulik}},
  \bibinfo {author} {\bibfnamefont {N.}~\bibnamefont {Ivanova}}, \bibinfo
  {author} {\bibfnamefont {F.~A.}\ \bibnamefont {Rasio}}, \ and\ \bibinfo
  {author} {\bibfnamefont {R.}~\bibnamefont {O'Shaughnessy}},\ }\href@noop {}
  {\bibfield  {journal} {\bibinfo  {journal} {The Astrophysical Journal}\
  }\textbf {\bibinfo {volume} {676}},\ \bibinfo {pages} {1162} (\bibinfo {year}
  {2008})}\BibitemShut {NoStop}%
\bibitem [{\citenamefont {Downing}\ \emph {et~al.}(2010)\citenamefont
  {Downing}, \citenamefont {Benacquista}, \citenamefont {Giersz},\ and\
  \citenamefont {Spurzem}}]{downing2010compact}%
  \BibitemOpen
  \bibfield  {author} {\bibinfo {author} {\bibfnamefont {J.}~\bibnamefont
  {Downing}}, \bibinfo {author} {\bibfnamefont {M.}~\bibnamefont
  {Benacquista}}, \bibinfo {author} {\bibfnamefont {M.}~\bibnamefont {Giersz}},
  \ and\ \bibinfo {author} {\bibfnamefont {R.}~\bibnamefont {Spurzem}},\
  }\href@noop {} {\bibfield  {journal} {\bibinfo  {journal} {Monthly Notices of
  the Royal Astronomical Society}\ }\textbf {\bibinfo {volume} {407}},\
  \bibinfo {pages} {1946} (\bibinfo {year} {2010})}\BibitemShut {NoStop}%
\bibitem [{\citenamefont {Downing}\ \emph {et~al.}(2011)\citenamefont
  {Downing}, \citenamefont {Benacquista}, \citenamefont {Giersz},\ and\
  \citenamefont {Spurzem}}]{downing2011compact}%
  \BibitemOpen
  \bibfield  {author} {\bibinfo {author} {\bibfnamefont {J.}~\bibnamefont
  {Downing}}, \bibinfo {author} {\bibfnamefont {M.}~\bibnamefont
  {Benacquista}}, \bibinfo {author} {\bibfnamefont {M.}~\bibnamefont {Giersz}},
  \ and\ \bibinfo {author} {\bibfnamefont {R.}~\bibnamefont {Spurzem}},\
  }\href@noop {} {\bibfield  {journal} {\bibinfo  {journal} {Monthly Notices of
  the Royal Astronomical Society}\ }\textbf {\bibinfo {volume} {416}},\
  \bibinfo {pages} {133} (\bibinfo {year} {2011})}\BibitemShut {NoStop}%
\bibitem [{\citenamefont {Samsing}\ \emph {et~al.}(2014)\citenamefont
  {Samsing}, \citenamefont {MacLeod},\ and\ \citenamefont
  {Ramirez-Ruiz}}]{Samsing:2013kua}%
  \BibitemOpen
  \bibfield  {author} {\bibinfo {author} {\bibfnamefont {J.}~\bibnamefont
  {Samsing}}, \bibinfo {author} {\bibfnamefont {M.}~\bibnamefont {MacLeod}}, \
  and\ \bibinfo {author} {\bibfnamefont {E.}~\bibnamefont {Ramirez-Ruiz}},\
  }\href {\doibase 10.1088/0004-637X/784/1/71} {\bibfield  {journal} {\bibinfo
  {journal} {Astrophys. J.}\ }\textbf {\bibinfo {volume} {784}},\ \bibinfo
  {pages} {71} (\bibinfo {year} {2014})}\BibitemShut {NoStop}%
\bibitem [{\citenamefont {Rodriguez}\ \emph {et~al.}(2015)\citenamefont
  {Rodriguez}, \citenamefont {Morscher}, \citenamefont {Pattabiraman},
  \citenamefont {Chatterjee}, \citenamefont {Haster},\ and\ \citenamefont
  {Rasio}}]{PhysRevLett.115.051101}%
  \BibitemOpen
  \bibfield  {author} {\bibinfo {author} {\bibfnamefont {C.~L.}\ \bibnamefont
  {Rodriguez}}, \bibinfo {author} {\bibfnamefont {M.}~\bibnamefont {Morscher}},
  \bibinfo {author} {\bibfnamefont {B.}~\bibnamefont {Pattabiraman}}, \bibinfo
  {author} {\bibfnamefont {S.}~\bibnamefont {Chatterjee}}, \bibinfo {author}
  {\bibfnamefont {C.-J.}\ \bibnamefont {Haster}}, \ and\ \bibinfo {author}
  {\bibfnamefont {F.~A.}\ \bibnamefont {Rasio}},\ }\href {\doibase
  10.1103/PhysRevLett.115.051101} {\bibfield  {journal} {\bibinfo  {journal}
  {Phys. Rev. Lett.}\ }\textbf {\bibinfo {volume} {115}},\ \bibinfo {pages}
  {051101} (\bibinfo {year} {2015})}\BibitemShut {NoStop}%
\bibitem [{\citenamefont {Rodriguez}\ \emph
  {et~al.}(2016{\natexlab{a}})\citenamefont {Rodriguez}, \citenamefont
  {Chatterjee},\ and\ \citenamefont {Rasio}}]{rodriguez2016binary}%
  \BibitemOpen
  \bibfield  {author} {\bibinfo {author} {\bibfnamefont {C.~L.}\ \bibnamefont
  {Rodriguez}}, \bibinfo {author} {\bibfnamefont {S.}~\bibnamefont
  {Chatterjee}}, \ and\ \bibinfo {author} {\bibfnamefont {F.~A.}\ \bibnamefont
  {Rasio}},\ }\href@noop {} {\bibfield  {journal} {\bibinfo  {journal}
  {Physical Review D}\ }\textbf {\bibinfo {volume} {93}},\ \bibinfo {pages}
  {084029} (\bibinfo {year} {2016}{\natexlab{a}})}\BibitemShut {NoStop}%
\bibitem [{\citenamefont {Askar}\ \emph {et~al.}(2016)\citenamefont {Askar},
  \citenamefont {Szkudlarek}, \citenamefont {Gondek-Rosi{\'n}ska},
  \citenamefont {Giersz},\ and\ \citenamefont {Bulik}}]{askar2016mocca}%
  \BibitemOpen
  \bibfield  {author} {\bibinfo {author} {\bibfnamefont {A.}~\bibnamefont
  {Askar}}, \bibinfo {author} {\bibfnamefont {M.}~\bibnamefont {Szkudlarek}},
  \bibinfo {author} {\bibfnamefont {D.}~\bibnamefont {Gondek-Rosi{\'n}ska}},
  \bibinfo {author} {\bibfnamefont {M.}~\bibnamefont {Giersz}}, \ and\ \bibinfo
  {author} {\bibfnamefont {T.}~\bibnamefont {Bulik}},\ }\href@noop {}
  {\bibfield  {journal} {\bibinfo  {journal} {Monthly Notices of the Royal
  Astronomical Society: Letters}\ }\textbf {\bibinfo {volume} {464}},\ \bibinfo
  {pages} {L36} (\bibinfo {year} {2016})}\BibitemShut {NoStop}%
\bibitem [{\citenamefont {Antonini}\ and\ \citenamefont
  {Rasio}(2016)}]{antonini2016merging}%
  \BibitemOpen
  \bibfield  {author} {\bibinfo {author} {\bibfnamefont {F.}~\bibnamefont
  {Antonini}}\ and\ \bibinfo {author} {\bibfnamefont {F.~A.}\ \bibnamefont
  {Rasio}},\ }\href@noop {} {\bibfield  {journal} {\bibinfo  {journal} {The
  Astrophysical Journal}\ }\textbf {\bibinfo {volume} {831}},\ \bibinfo {pages}
  {187} (\bibinfo {year} {2016})}\BibitemShut {NoStop}%
\bibitem [{\citenamefont {Petrovich}\ and\ \citenamefont
  {Antonini}(2017)}]{petrovich2017greatly}%
  \BibitemOpen
  \bibfield  {author} {\bibinfo {author} {\bibfnamefont {C.}~\bibnamefont
  {Petrovich}}\ and\ \bibinfo {author} {\bibfnamefont {F.}~\bibnamefont
  {Antonini}},\ }\href@noop {} {\bibfield  {journal} {\bibinfo  {journal} {The
  Astrophysical Journal}\ }\textbf {\bibinfo {volume} {846}},\ \bibinfo {pages}
  {146} (\bibinfo {year} {2017})}\BibitemShut {NoStop}%
\bibitem [{\citenamefont {Ziosi}\ \emph {et~al.}(2014)\citenamefont {Ziosi},
  \citenamefont {Mapelli}, \citenamefont {Branchesi},\ and\ \citenamefont
  {Tormen}}]{ziosi2014dynamics}%
  \BibitemOpen
  \bibfield  {author} {\bibinfo {author} {\bibfnamefont {B.~M.}\ \bibnamefont
  {Ziosi}}, \bibinfo {author} {\bibfnamefont {M.}~\bibnamefont {Mapelli}},
  \bibinfo {author} {\bibfnamefont {M.}~\bibnamefont {Branchesi}}, \ and\
  \bibinfo {author} {\bibfnamefont {G.}~\bibnamefont {Tormen}},\ }\href@noop {}
  {\bibfield  {journal} {\bibinfo  {journal} {Monthly Notices of the Royal
  Astronomical Society}\ }\textbf {\bibinfo {volume} {441}},\ \bibinfo {pages}
  {3703} (\bibinfo {year} {2014})}\BibitemShut {NoStop}%
\bibitem [{\citenamefont {Mapelli}(2016)}]{mapelli2016massive}%
  \BibitemOpen
  \bibfield  {author} {\bibinfo {author} {\bibfnamefont {M.}~\bibnamefont
  {Mapelli}},\ }\href@noop {} {\bibfield  {journal} {\bibinfo  {journal}
  {Monthly Notices of the Royal Astronomical Society}\ }\textbf {\bibinfo
  {volume} {459}},\ \bibinfo {pages} {3432} (\bibinfo {year}
  {2016})}\BibitemShut {NoStop}%
\bibitem [{\citenamefont {Banerjee}(2017)}]{banerjee2017stellar}%
  \BibitemOpen
  \bibfield  {author} {\bibinfo {author} {\bibfnamefont {S.}~\bibnamefont
  {Banerjee}},\ }\href@noop {} {\bibfield  {journal} {\bibinfo  {journal}
  {Monthly Notices of the Royal Astronomical Society}\ }\textbf {\bibinfo
  {volume} {467}},\ \bibinfo {pages} {524} (\bibinfo {year}
  {2017})}\BibitemShut {NoStop}%
\bibitem [{\citenamefont {Chatterjee}\ \emph {et~al.}(2017)\citenamefont
  {Chatterjee}, \citenamefont {Rodriguez}, \citenamefont {Kalogera},\ and\
  \citenamefont {Rasio}}]{chatterjee2017dynamical}%
  \BibitemOpen
  \bibfield  {author} {\bibinfo {author} {\bibfnamefont {S.}~\bibnamefont
  {Chatterjee}}, \bibinfo {author} {\bibfnamefont {C.~L.}\ \bibnamefont
  {Rodriguez}}, \bibinfo {author} {\bibfnamefont {V.}~\bibnamefont {Kalogera}},
  \ and\ \bibinfo {author} {\bibfnamefont {F.~A.}\ \bibnamefont {Rasio}},\
  }\href@noop {} {\bibfield  {journal} {\bibinfo  {journal} {The Astrophysical
  Journal Letters}\ }\textbf {\bibinfo {volume} {836}},\ \bibinfo {pages} {L26}
  (\bibinfo {year} {2017})}\BibitemShut {NoStop}%
\bibitem [{\citenamefont {Nelemans}\ \emph {et~al.}(2001)\citenamefont
  {Nelemans}, \citenamefont {Yungelson},\ and\ \citenamefont
  {Zwart}}]{nelemans2001gravitational}%
  \BibitemOpen
  \bibfield  {author} {\bibinfo {author} {\bibfnamefont {G.}~\bibnamefont
  {Nelemans}}, \bibinfo {author} {\bibfnamefont {L.}~\bibnamefont {Yungelson}},
  \ and\ \bibinfo {author} {\bibfnamefont {S.~P.}\ \bibnamefont {Zwart}},\
  }\href@noop {} {\bibfield  {journal} {\bibinfo  {journal} {Astronomy \&
  Astrophysics}\ }\textbf {\bibinfo {volume} {375}},\ \bibinfo {pages} {890}
  (\bibinfo {year} {2001})}\BibitemShut {NoStop}%
\bibitem [{\citenamefont {Belczynski}\ \emph {et~al.}(2002)\citenamefont
  {Belczynski}, \citenamefont {Kalogera},\ and\ \citenamefont
  {Bulik}}]{belczynski2002comprehensive}%
  \BibitemOpen
  \bibfield  {author} {\bibinfo {author} {\bibfnamefont {K.}~\bibnamefont
  {Belczynski}}, \bibinfo {author} {\bibfnamefont {V.}~\bibnamefont
  {Kalogera}}, \ and\ \bibinfo {author} {\bibfnamefont {T.}~\bibnamefont
  {Bulik}},\ }\href@noop {} {\bibfield  {journal} {\bibinfo  {journal} {The
  Astrophysical Journal}\ }\textbf {\bibinfo {volume} {572}},\ \bibinfo {pages}
  {407} (\bibinfo {year} {2002})}\BibitemShut {NoStop}%
\bibitem [{\citenamefont {Voss}\ and\ \citenamefont
  {Tauris}(2003)}]{voss2003galactic}%
  \BibitemOpen
  \bibfield  {author} {\bibinfo {author} {\bibfnamefont {R.}~\bibnamefont
  {Voss}}\ and\ \bibinfo {author} {\bibfnamefont {T.~M.}\ \bibnamefont
  {Tauris}},\ }\href@noop {} {\bibfield  {journal} {\bibinfo  {journal}
  {Monthly Notices of the Royal Astronomical Society}\ }\textbf {\bibinfo
  {volume} {342}},\ \bibinfo {pages} {1169} (\bibinfo {year}
  {2003})}\BibitemShut {NoStop}%
\bibitem [{\citenamefont {Belczynski}\ \emph {et~al.}(2007)\citenamefont
  {Belczynski}, \citenamefont {Taam}, \citenamefont {Kalogera}, \citenamefont
  {Rasio},\ and\ \citenamefont {Bulik}}]{belczynski2007rarity}%
  \BibitemOpen
  \bibfield  {author} {\bibinfo {author} {\bibfnamefont {K.}~\bibnamefont
  {Belczynski}}, \bibinfo {author} {\bibfnamefont {R.~E.}\ \bibnamefont
  {Taam}}, \bibinfo {author} {\bibfnamefont {V.}~\bibnamefont {Kalogera}},
  \bibinfo {author} {\bibfnamefont {F.~A.}\ \bibnamefont {Rasio}}, \ and\
  \bibinfo {author} {\bibfnamefont {T.}~\bibnamefont {Bulik}},\ }\href@noop {}
  {\bibfield  {journal} {\bibinfo  {journal} {The Astrophysical Journal}\
  }\textbf {\bibinfo {volume} {662}},\ \bibinfo {pages} {504} (\bibinfo {year}
  {2007})}\BibitemShut {NoStop}%
\bibitem [{\citenamefont {Belczynski}\ \emph {et~al.}(2008)\citenamefont
  {Belczynski}, \citenamefont {Kalogera}, \citenamefont {Rasio}, \citenamefont
  {Taam}, \citenamefont {Zezas}, \citenamefont {Bulik}, \citenamefont
  {Maccarone},\ and\ \citenamefont {Ivanova}}]{belczynski2008compact}%
  \BibitemOpen
  \bibfield  {author} {\bibinfo {author} {\bibfnamefont {K.}~\bibnamefont
  {Belczynski}}, \bibinfo {author} {\bibfnamefont {V.}~\bibnamefont
  {Kalogera}}, \bibinfo {author} {\bibfnamefont {F.~A.}\ \bibnamefont {Rasio}},
  \bibinfo {author} {\bibfnamefont {R.~E.}\ \bibnamefont {Taam}}, \bibinfo
  {author} {\bibfnamefont {A.}~\bibnamefont {Zezas}}, \bibinfo {author}
  {\bibfnamefont {T.}~\bibnamefont {Bulik}}, \bibinfo {author} {\bibfnamefont
  {T.~J.}\ \bibnamefont {Maccarone}}, \ and\ \bibinfo {author} {\bibfnamefont
  {N.}~\bibnamefont {Ivanova}},\ }\href@noop {} {\bibfield  {journal} {\bibinfo
   {journal} {The Astrophysical Journal Supplement Series}\ }\textbf {\bibinfo
  {volume} {174}},\ \bibinfo {pages} {223} (\bibinfo {year}
  {2008})}\BibitemShut {NoStop}%
\bibitem [{\citenamefont {Dominik}\ \emph {et~al.}(2013)\citenamefont
  {Dominik}, \citenamefont {Belczynski}, \citenamefont {Fryer}, \citenamefont
  {Holz}, \citenamefont {Berti}, \citenamefont {Bulik}, \citenamefont
  {Mandel},\ and\ \citenamefont {O'Shaughnessy}}]{dominik2013double}%
  \BibitemOpen
  \bibfield  {author} {\bibinfo {author} {\bibfnamefont {M.}~\bibnamefont
  {Dominik}}, \bibinfo {author} {\bibfnamefont {K.}~\bibnamefont {Belczynski}},
  \bibinfo {author} {\bibfnamefont {C.}~\bibnamefont {Fryer}}, \bibinfo
  {author} {\bibfnamefont {D.~E.}\ \bibnamefont {Holz}}, \bibinfo {author}
  {\bibfnamefont {E.}~\bibnamefont {Berti}}, \bibinfo {author} {\bibfnamefont
  {T.}~\bibnamefont {Bulik}}, \bibinfo {author} {\bibfnamefont
  {I.}~\bibnamefont {Mandel}}, \ and\ \bibinfo {author} {\bibfnamefont
  {R.}~\bibnamefont {O'Shaughnessy}},\ }\href@noop {} {\bibfield  {journal}
  {\bibinfo  {journal} {The Astrophysical Journal}\ }\textbf {\bibinfo {volume}
  {779}},\ \bibinfo {pages} {72} (\bibinfo {year} {2013})}\BibitemShut
  {NoStop}%
\bibitem [{\citenamefont {Belczynski}\ \emph {et~al.}(2014)\citenamefont
  {Belczynski}, \citenamefont {Buonanno}, \citenamefont {Cantiello},
  \citenamefont {Fryer}, \citenamefont {Holz}, \citenamefont {Mandel},
  \citenamefont {Miller},\ and\ \citenamefont
  {Walczak}}]{belczynski2014formation}%
  \BibitemOpen
  \bibfield  {author} {\bibinfo {author} {\bibfnamefont {K.}~\bibnamefont
  {Belczynski}}, \bibinfo {author} {\bibfnamefont {A.}~\bibnamefont
  {Buonanno}}, \bibinfo {author} {\bibfnamefont {M.}~\bibnamefont {Cantiello}},
  \bibinfo {author} {\bibfnamefont {C.~L.}\ \bibnamefont {Fryer}}, \bibinfo
  {author} {\bibfnamefont {D.~E.}\ \bibnamefont {Holz}}, \bibinfo {author}
  {\bibfnamefont {I.}~\bibnamefont {Mandel}}, \bibinfo {author} {\bibfnamefont
  {M.~C.}\ \bibnamefont {Miller}}, \ and\ \bibinfo {author} {\bibfnamefont
  {M.}~\bibnamefont {Walczak}},\ }\href@noop {} {\bibfield  {journal} {\bibinfo
   {journal} {The Astrophysical Journal}\ }\textbf {\bibinfo {volume} {789}},\
  \bibinfo {pages} {120} (\bibinfo {year} {2014})}\BibitemShut {NoStop}%
\bibitem [{\citenamefont {Mennekens}\ and\ \citenamefont
  {Vanbeveren}(2014)}]{mennekens2014massive}%
  \BibitemOpen
  \bibfield  {author} {\bibinfo {author} {\bibfnamefont {N.}~\bibnamefont
  {Mennekens}}\ and\ \bibinfo {author} {\bibfnamefont {D.}~\bibnamefont
  {Vanbeveren}},\ }\href@noop {} {\bibfield  {journal} {\bibinfo  {journal}
  {Astronomy \& Astrophysics}\ }\textbf {\bibinfo {volume} {564}},\ \bibinfo
  {pages} {A134} (\bibinfo {year} {2014})}\BibitemShut {NoStop}%
\bibitem [{\citenamefont {Spera}\ \emph {et~al.}(2015)\citenamefont {Spera},
  \citenamefont {Mapelli},\ and\ \citenamefont {Bressan}}]{spera2015mass}%
  \BibitemOpen
  \bibfield  {author} {\bibinfo {author} {\bibfnamefont {M.}~\bibnamefont
  {Spera}}, \bibinfo {author} {\bibfnamefont {M.}~\bibnamefont {Mapelli}}, \
  and\ \bibinfo {author} {\bibfnamefont {A.}~\bibnamefont {Bressan}},\
  }\href@noop {} {\bibfield  {journal} {\bibinfo  {journal} {Monthly Notices of
  the Royal Astronomical Society}\ }\textbf {\bibinfo {volume} {451}},\
  \bibinfo {pages} {4086} (\bibinfo {year} {2015})}\BibitemShut {NoStop}%
\bibitem [{\citenamefont {Eldridge}\ and\ \citenamefont
  {Stanway}(2016)}]{eldridge2016bpass}%
  \BibitemOpen
  \bibfield  {author} {\bibinfo {author} {\bibfnamefont {J.}~\bibnamefont
  {Eldridge}}\ and\ \bibinfo {author} {\bibfnamefont {E.}~\bibnamefont
  {Stanway}},\ }\href@noop {} {\bibfield  {journal} {\bibinfo  {journal}
  {Monthly Notices of the Royal Astronomical Society}\ }\textbf {\bibinfo
  {volume} {462}},\ \bibinfo {pages} {3302} (\bibinfo {year}
  {2016})}\BibitemShut {NoStop}%
\bibitem [{\citenamefont {Stevenson}\ \emph {et~al.}(2017)\citenamefont
  {Stevenson}, \citenamefont {Vigna-G{\'o}mez}, \citenamefont {Mandel},
  \citenamefont {Barrett}, \citenamefont {Neijssel}, \citenamefont {Perkins},\
  and\ \citenamefont {de~Mink}}]{stevenson2017formation}%
  \BibitemOpen
  \bibfield  {author} {\bibinfo {author} {\bibfnamefont {S.}~\bibnamefont
  {Stevenson}}, \bibinfo {author} {\bibfnamefont {A.}~\bibnamefont
  {Vigna-G{\'o}mez}}, \bibinfo {author} {\bibfnamefont {I.}~\bibnamefont
  {Mandel}}, \bibinfo {author} {\bibfnamefont {J.~W.}\ \bibnamefont {Barrett}},
  \bibinfo {author} {\bibfnamefont {C.~J.}\ \bibnamefont {Neijssel}}, \bibinfo
  {author} {\bibfnamefont {D.}~\bibnamefont {Perkins}}, \ and\ \bibinfo
  {author} {\bibfnamefont {S.~E.}\ \bibnamefont {de~Mink}},\ }\href@noop {}
  {\bibfield  {journal} {\bibinfo  {journal} {Nature communications}\ }\textbf
  {\bibinfo {volume} {8}},\ \bibinfo {pages} {14906} (\bibinfo {year}
  {2017})}\BibitemShut {NoStop}%
\bibitem [{\citenamefont {Mapelli}\ \emph {et~al.}(2017)\citenamefont
  {Mapelli}, \citenamefont {Giacobbo}, \citenamefont {Ripamonti},\ and\
  \citenamefont {Spera}}]{mapelli2017cosmic}%
  \BibitemOpen
  \bibfield  {author} {\bibinfo {author} {\bibfnamefont {M.}~\bibnamefont
  {Mapelli}}, \bibinfo {author} {\bibfnamefont {N.}~\bibnamefont {Giacobbo}},
  \bibinfo {author} {\bibfnamefont {E.}~\bibnamefont {Ripamonti}}, \ and\
  \bibinfo {author} {\bibfnamefont {M.}~\bibnamefont {Spera}},\ }\href@noop {}
  {\bibfield  {journal} {\bibinfo  {journal} {Monthly Notices of the Royal
  Astronomical Society}\ }\textbf {\bibinfo {volume} {472}},\ \bibinfo {pages}
  {2422} (\bibinfo {year} {2017})}\BibitemShut {NoStop}%
\bibitem [{\citenamefont {Giacobbo}\ \emph {et~al.}(2017)\citenamefont
  {Giacobbo}, \citenamefont {Mapelli},\ and\ \citenamefont
  {Spera}}]{giacobbo2017merging}%
  \BibitemOpen
  \bibfield  {author} {\bibinfo {author} {\bibfnamefont {N.}~\bibnamefont
  {Giacobbo}}, \bibinfo {author} {\bibfnamefont {M.}~\bibnamefont {Mapelli}}, \
  and\ \bibinfo {author} {\bibfnamefont {M.}~\bibnamefont {Spera}},\
  }\href@noop {} {\bibfield  {journal} {\bibinfo  {journal} {Monthly Notices of
  the Royal Astronomical Society}\ }\textbf {\bibinfo {volume} {474}},\
  \bibinfo {pages} {2959} (\bibinfo {year} {2017})}\BibitemShut {NoStop}%
\bibitem [{\citenamefont {Mapelli}\ and\ \citenamefont
  {Giacobbo}(2018)}]{mapelli2018cosmic}%
  \BibitemOpen
  \bibfield  {author} {\bibinfo {author} {\bibfnamefont {M.}~\bibnamefont
  {Mapelli}}\ and\ \bibinfo {author} {\bibfnamefont {N.}~\bibnamefont
  {Giacobbo}},\ }\href@noop {} {\bibfield  {journal} {\bibinfo  {journal}
  {Monthly Notices of the Royal Astronomical Society}\ } (\bibinfo {year}
  {2018})}\BibitemShut {NoStop}%
\bibitem [{\citenamefont {Kruckow}\ \emph {et~al.}(2018)\citenamefont
  {Kruckow}, \citenamefont {Tauris}, \citenamefont {Langer}, \citenamefont
  {Kramer},\ and\ \citenamefont {Izzard}}]{kruckow2018progenitors}%
  \BibitemOpen
  \bibfield  {author} {\bibinfo {author} {\bibfnamefont {M.~U.}\ \bibnamefont
  {Kruckow}}, \bibinfo {author} {\bibfnamefont {T.~M.}\ \bibnamefont {Tauris}},
  \bibinfo {author} {\bibfnamefont {N.}~\bibnamefont {Langer}}, \bibinfo
  {author} {\bibfnamefont {M.}~\bibnamefont {Kramer}}, \ and\ \bibinfo {author}
  {\bibfnamefont {R.~G.}\ \bibnamefont {Izzard}},\ }\href {\doibase
  10.1093/mnras/sty2190} {\bibfield  {journal} {\bibinfo  {journal} {Monthly
  Notices of the Royal Astronomical Society}\ }\textbf {\bibinfo {volume}
  {481}},\ \bibinfo {pages} {1908} (\bibinfo {year} {2018})}\BibitemShut
  {NoStop}%
\bibitem [{\citenamefont {Giacobbo}\ and\ \citenamefont
  {Mapelli}(2018)}]{giacobbo2018progenitors}%
  \BibitemOpen
  \bibfield  {author} {\bibinfo {author} {\bibfnamefont {N.}~\bibnamefont
  {Giacobbo}}\ and\ \bibinfo {author} {\bibfnamefont {M.}~\bibnamefont
  {Mapelli}},\ }\href {\doibase 10.1093/mnras/sty1999} {\bibfield  {journal}
  {\bibinfo  {journal} {Monthly Notices of the Royal Astronomical Society}\
  }\textbf {\bibinfo {volume} {480}},\ \bibinfo {pages} {2011} (\bibinfo {year}
  {2018})}\BibitemShut {NoStop}%
\bibitem [{\citenamefont {Marchant}\ \emph {et~al.}(2016)\citenamefont
  {Marchant}, \citenamefont {Langer}, \citenamefont {Podsiadlowski},
  \citenamefont {Tauris},\ and\ \citenamefont {Moriya}}]{marchant2016new}%
  \BibitemOpen
  \bibfield  {author} {\bibinfo {author} {\bibfnamefont {P.}~\bibnamefont
  {Marchant}}, \bibinfo {author} {\bibfnamefont {N.}~\bibnamefont {Langer}},
  \bibinfo {author} {\bibfnamefont {P.}~\bibnamefont {Podsiadlowski}}, \bibinfo
  {author} {\bibfnamefont {T.~M.}\ \bibnamefont {Tauris}}, \ and\ \bibinfo
  {author} {\bibfnamefont {T.~J.}\ \bibnamefont {Moriya}},\ }\href@noop {}
  {\bibfield  {journal} {\bibinfo  {journal} {Astronomy \& Astrophysics}\
  }\textbf {\bibinfo {volume} {588}},\ \bibinfo {pages} {A50} (\bibinfo {year}
  {2016})}\BibitemShut {NoStop}%
\bibitem [{\citenamefont {De~Mink}\ and\ \citenamefont
  {Mandel}(2016)}]{de2016chemically}%
  \BibitemOpen
  \bibfield  {author} {\bibinfo {author} {\bibfnamefont {S.}~\bibnamefont
  {De~Mink}}\ and\ \bibinfo {author} {\bibfnamefont {I.}~\bibnamefont
  {Mandel}},\ }\href@noop {} {\bibfield  {journal} {\bibinfo  {journal}
  {Monthly Notices of the Royal Astronomical Society}\ }\textbf {\bibinfo
  {volume} {460}},\ \bibinfo {pages} {3545} (\bibinfo {year}
  {2016})}\BibitemShut {NoStop}%
\bibitem [{\citenamefont {Mandel}\ and\ \citenamefont
  {de~Mink}(2016)}]{mandel2016merging}%
  \BibitemOpen
  \bibfield  {author} {\bibinfo {author} {\bibfnamefont {I.}~\bibnamefont
  {Mandel}}\ and\ \bibinfo {author} {\bibfnamefont {S.~E.}\ \bibnamefont
  {de~Mink}},\ }\href@noop {} {\bibfield  {journal} {\bibinfo  {journal}
  {Monthly Notices of the Royal Astronomical Society}\ }\textbf {\bibinfo
  {volume} {458}},\ \bibinfo {pages} {2634} (\bibinfo {year}
  {2016})}\BibitemShut {NoStop}%
\bibitem [{\citenamefont {Antonini}\ and\ \citenamefont
  {Perets}(2012)}]{antonini2012secular}%
  \BibitemOpen
  \bibfield  {author} {\bibinfo {author} {\bibfnamefont {F.}~\bibnamefont
  {Antonini}}\ and\ \bibinfo {author} {\bibfnamefont {H.~B.}\ \bibnamefont
  {Perets}},\ }\href@noop {} {\bibfield  {journal} {\bibinfo  {journal} {The
  Astrophysical Journal}\ }\textbf {\bibinfo {volume} {757}},\ \bibinfo {pages}
  {27} (\bibinfo {year} {2012})}\BibitemShut {NoStop}%
\bibitem [{\citenamefont {McKernan}\ \emph {et~al.}(2012)\citenamefont
  {McKernan}, \citenamefont {Ford}, \citenamefont {Lyra},\ and\ \citenamefont
  {Perets}}]{mckernan2012intermediate}%
  \BibitemOpen
  \bibfield  {author} {\bibinfo {author} {\bibfnamefont {B.}~\bibnamefont
  {McKernan}}, \bibinfo {author} {\bibfnamefont {K.}~\bibnamefont {Ford}},
  \bibinfo {author} {\bibfnamefont {W.}~\bibnamefont {Lyra}}, \ and\ \bibinfo
  {author} {\bibfnamefont {H.}~\bibnamefont {Perets}},\ }\href@noop {}
  {\bibfield  {journal} {\bibinfo  {journal} {Monthly Notices of the Royal
  Astronomical Society}\ }\textbf {\bibinfo {volume} {425}},\ \bibinfo {pages}
  {460} (\bibinfo {year} {2012})}\BibitemShut {NoStop}%
\bibitem [{\citenamefont {Stone}\ \emph {et~al.}(2016)\citenamefont {Stone},
  \citenamefont {Metzger},\ and\ \citenamefont {Haiman}}]{stone2016assisted}%
  \BibitemOpen
  \bibfield  {author} {\bibinfo {author} {\bibfnamefont {N.~C.}\ \bibnamefont
  {Stone}}, \bibinfo {author} {\bibfnamefont {B.~D.}\ \bibnamefont {Metzger}},
  \ and\ \bibinfo {author} {\bibfnamefont {Z.}~\bibnamefont {Haiman}},\
  }\href@noop {} {\bibfield  {journal} {\bibinfo  {journal} {Monthly Notices of
  the Royal Astronomical Society}\ }\textbf {\bibinfo {volume} {464}},\
  \bibinfo {pages} {946} (\bibinfo {year} {2016})}\BibitemShut {NoStop}%
\bibitem [{\citenamefont {Bartos}\ \emph {et~al.}(2017)\citenamefont {Bartos},
  \citenamefont {Kocsis}, \citenamefont {Haiman},\ and\ \citenamefont
  {M{\'a}rka}}]{bartos2017rapid}%
  \BibitemOpen
  \bibfield  {author} {\bibinfo {author} {\bibfnamefont {I.}~\bibnamefont
  {Bartos}}, \bibinfo {author} {\bibfnamefont {B.}~\bibnamefont {Kocsis}},
  \bibinfo {author} {\bibfnamefont {Z.}~\bibnamefont {Haiman}}, \ and\ \bibinfo
  {author} {\bibfnamefont {S.}~\bibnamefont {M{\'a}rka}},\ }\href@noop {}
  {\bibfield  {journal} {\bibinfo  {journal} {The Astrophysical Journal}\
  }\textbf {\bibinfo {volume} {835}},\ \bibinfo {pages} {165} (\bibinfo {year}
  {2017})}\BibitemShut {NoStop}%
\bibitem [{\citenamefont {Antonini}\ \emph {et~al.}(2014)\citenamefont
  {Antonini}, \citenamefont {Murray},\ and\ \citenamefont
  {Mikkola}}]{antonini2014black}%
  \BibitemOpen
  \bibfield  {author} {\bibinfo {author} {\bibfnamefont {F.}~\bibnamefont
  {Antonini}}, \bibinfo {author} {\bibfnamefont {N.}~\bibnamefont {Murray}}, \
  and\ \bibinfo {author} {\bibfnamefont {S.}~\bibnamefont {Mikkola}},\
  }\href@noop {} {\bibfield  {journal} {\bibinfo  {journal} {The Astrophysical
  Journal}\ }\textbf {\bibinfo {volume} {781}},\ \bibinfo {pages} {45}
  (\bibinfo {year} {2014})}\BibitemShut {NoStop}%
\bibitem [{\citenamefont {Kimpson}\ \emph {et~al.}(2016)\citenamefont
  {Kimpson}, \citenamefont {Spera}, \citenamefont {Mapelli},\ and\
  \citenamefont {Ziosi}}]{kimpson2016hierarchical}%
  \BibitemOpen
  \bibfield  {author} {\bibinfo {author} {\bibfnamefont {T.~O.}\ \bibnamefont
  {Kimpson}}, \bibinfo {author} {\bibfnamefont {M.}~\bibnamefont {Spera}},
  \bibinfo {author} {\bibfnamefont {M.}~\bibnamefont {Mapelli}}, \ and\
  \bibinfo {author} {\bibfnamefont {B.~M.}\ \bibnamefont {Ziosi}},\ }\href@noop
  {} {\bibfield  {journal} {\bibinfo  {journal} {Monthly Notices of the Royal
  Astronomical Society}\ }\textbf {\bibinfo {volume} {463}},\ \bibinfo {pages}
  {2443} (\bibinfo {year} {2016})}\BibitemShut {NoStop}%
\bibitem [{\citenamefont {Antonini}\ \emph {et~al.}(2017)\citenamefont
  {Antonini}, \citenamefont {Toonen},\ and\ \citenamefont
  {Hamers}}]{antonini2017binary}%
  \BibitemOpen
  \bibfield  {author} {\bibinfo {author} {\bibfnamefont {F.}~\bibnamefont
  {Antonini}}, \bibinfo {author} {\bibfnamefont {S.}~\bibnamefont {Toonen}}, \
  and\ \bibinfo {author} {\bibfnamefont {A.~S.}\ \bibnamefont {Hamers}},\
  }\href@noop {} {\bibfield  {journal} {\bibinfo  {journal} {The Astrophysical
  Journal}\ }\textbf {\bibinfo {volume} {841}},\ \bibinfo {pages} {77}
  (\bibinfo {year} {2017})}\BibitemShut {NoStop}%
\bibitem [{\citenamefont {Liu}\ and\ \citenamefont {Lai}(2018)}]{liu2018black}%
  \BibitemOpen
  \bibfield  {author} {\bibinfo {author} {\bibfnamefont {B.}~\bibnamefont
  {Liu}}\ and\ \bibinfo {author} {\bibfnamefont {D.}~\bibnamefont {Lai}},\
  }\href {\doibase 10.3847/1538-4357/aad09f} {\bibfield  {journal} {\bibinfo
  {journal} {The Astrophysical Journal}\ }\textbf {\bibinfo {volume} {863}},\
  \bibinfo {pages} {68} (\bibinfo {year} {2018})}\BibitemShut {NoStop}%
\bibitem [{\citenamefont {Hamers}\ \emph {et~al.}(2015)\citenamefont {Hamers},
  \citenamefont {Perets}, \citenamefont {Antonini},\ and\ \citenamefont
  {Zwart}}]{Hamers2015}%
  \BibitemOpen
  \bibfield  {author} {\bibinfo {author} {\bibfnamefont {A.~S.}\ \bibnamefont
  {Hamers}}, \bibinfo {author} {\bibfnamefont {H.~B.}\ \bibnamefont {Perets}},
  \bibinfo {author} {\bibfnamefont {F.}~\bibnamefont {Antonini}}, \ and\
  \bibinfo {author} {\bibfnamefont {S.~F.~P.}\ \bibnamefont {Zwart}},\ }\href
  {\doibase 10.1093/mnras/stv452} {\bibfield  {journal} {\bibinfo  {journal}
  {Monthly Notices of the Royal Astronomical Society}\ }\textbf {\bibinfo
  {volume} {449}},\ \bibinfo {pages} {4221} (\bibinfo {year}
  {2015})}\BibitemShut {NoStop}%
\bibitem [{\citenamefont {Vitale}\ \emph {et~al.}(2014)\citenamefont {Vitale},
  \citenamefont {Lynch}, \citenamefont {Veitch}, \citenamefont {Raymond},\ and\
  \citenamefont {Sturani}}]{Vitale:2014mka}%
  \BibitemOpen
  \bibfield  {author} {\bibinfo {author} {\bibfnamefont {S.}~\bibnamefont
  {Vitale}}, \bibinfo {author} {\bibfnamefont {R.}~\bibnamefont {Lynch}},
  \bibinfo {author} {\bibfnamefont {J.}~\bibnamefont {Veitch}}, \bibinfo
  {author} {\bibfnamefont {V.}~\bibnamefont {Raymond}}, \ and\ \bibinfo
  {author} {\bibfnamefont {R.}~\bibnamefont {Sturani}},\ }\href {\doibase
  10.1103/PhysRevLett.112.251101} {\bibfield  {journal} {\bibinfo  {journal}
  {Phys. Rev. Lett.}\ }\textbf {\bibinfo {volume} {112}},\ \bibinfo {pages}
  {251101} (\bibinfo {year} {2014})},\ \Eprint {http://arxiv.org/abs/1403.0129}
  {arXiv:1403.0129 [gr-qc]} \BibitemShut {NoStop}%
\bibitem [{\citenamefont {P\"urrer}\ \emph {et~al.}(2016)\citenamefont
  {P\"urrer}, \citenamefont {Hannam},\ and\ \citenamefont
  {Ohme}}]{Purrer:2015nkh}%
  \BibitemOpen
  \bibfield  {author} {\bibinfo {author} {\bibfnamefont {M.}~\bibnamefont
  {P\"urrer}}, \bibinfo {author} {\bibfnamefont {M.}~\bibnamefont {Hannam}}, \
  and\ \bibinfo {author} {\bibfnamefont {F.}~\bibnamefont {Ohme}},\ }\href
  {\doibase 10.1103/PhysRevD.93.084042} {\bibfield  {journal} {\bibinfo
  {journal} {Phys. Rev. D}\ }\textbf {\bibinfo {volume} {93}},\ \bibinfo
  {pages} {084042} (\bibinfo {year} {2016})},\ \Eprint
  {http://arxiv.org/abs/1512.04955} {arXiv:1512.04955 [gr-qc]} \BibitemShut
  {NoStop}%
\bibitem [{\citenamefont {Vitale}\ \emph {et~al.}(2017)\citenamefont {Vitale},
  \citenamefont {Lynch}, \citenamefont {Raymond}, \citenamefont {Sturani},
  \citenamefont {Veitch},\ and\ \citenamefont {Graff}}]{Vitale:2016avz}%
  \BibitemOpen
  \bibfield  {author} {\bibinfo {author} {\bibfnamefont {S.}~\bibnamefont
  {Vitale}}, \bibinfo {author} {\bibfnamefont {R.}~\bibnamefont {Lynch}},
  \bibinfo {author} {\bibfnamefont {V.}~\bibnamefont {Raymond}}, \bibinfo
  {author} {\bibfnamefont {R.}~\bibnamefont {Sturani}}, \bibinfo {author}
  {\bibfnamefont {J.}~\bibnamefont {Veitch}}, \ and\ \bibinfo {author}
  {\bibfnamefont {P.}~\bibnamefont {Graff}},\ }\href {\doibase
  10.1103/PhysRevD.95.064053} {\bibfield  {journal} {\bibinfo  {journal} {Phys.
  Rev. D}\ }\textbf {\bibinfo {volume} {95}},\ \bibinfo {pages} {064053}
  (\bibinfo {year} {2017})},\ \Eprint {http://arxiv.org/abs/1611.01122}
  {arXiv:1611.01122 [gr-qc]} \BibitemShut {NoStop}%
\bibitem [{\citenamefont {Apostolatos}\ \emph {et~al.}(1994)\citenamefont
  {Apostolatos}, \citenamefont {Cutler}, \citenamefont {Sussman},\ and\
  \citenamefont {Thorne}}]{Apostolatos:1994mx}%
  \BibitemOpen
  \bibfield  {author} {\bibinfo {author} {\bibfnamefont {T.~A.}\ \bibnamefont
  {Apostolatos}}, \bibinfo {author} {\bibfnamefont {C.}~\bibnamefont {Cutler}},
  \bibinfo {author} {\bibfnamefont {G.~J.}\ \bibnamefont {Sussman}}, \ and\
  \bibinfo {author} {\bibfnamefont {K.~S.}\ \bibnamefont {Thorne}},\ }\href
  {\doibase 10.1103/PhysRevD.49.6274} {\bibfield  {journal} {\bibinfo
  {journal} {Phys. Rev. D}\ }\textbf {\bibinfo {volume} {49}},\ \bibinfo
  {pages} {6274} (\bibinfo {year} {1994})}\BibitemShut {NoStop}%
\bibitem [{\citenamefont {Kidder}(1995)}]{Kidder:1995zr}%
  \BibitemOpen
  \bibfield  {author} {\bibinfo {author} {\bibfnamefont {L.~E.}\ \bibnamefont
  {Kidder}},\ }\href {\doibase 10.1103/PhysRevD.52.821} {\bibfield  {journal}
  {\bibinfo  {journal} {Phys. Rev. D}\ }\textbf {\bibinfo {volume} {52}},\
  \bibinfo {pages} {821} (\bibinfo {year} {1995})},\ \Eprint
  {http://arxiv.org/abs/gr-qc/9506022} {arXiv:gr-qc/9506022} \BibitemShut
  {NoStop}%
\bibitem [{\citenamefont {Racine}(2008)}]{Racine:2008qv}%
  \BibitemOpen
  \bibfield  {author} {\bibinfo {author} {\bibfnamefont {E.}~\bibnamefont
  {Racine}},\ }\href {\doibase 10.1103/PhysRevD.78.044021} {\bibfield
  {journal} {\bibinfo  {journal} {Phys. Rev. D}\ }\textbf {\bibinfo {volume}
  {78}},\ \bibinfo {pages} {044021} (\bibinfo {year} {2008})},\ \Eprint
  {http://arxiv.org/abs/0803.1820} {arXiv:0803.1820 [gr-qc]} \BibitemShut
  {NoStop}%
\bibitem [{\citenamefont {Ajith}\ \emph {et~al.}(2011)\citenamefont {Ajith}
  \emph {et~al.}}]{Ajith:2009bn}%
  \BibitemOpen
  \bibfield  {author} {\bibinfo {author} {\bibfnamefont {P.}~\bibnamefont
  {Ajith}} \emph {et~al.},\ }\href {\doibase 10.1103/PhysRevLett.106.241101}
  {\bibfield  {journal} {\bibinfo  {journal} {Phys. Rev. Lett.}\ }\textbf
  {\bibinfo {volume} {106}},\ \bibinfo {pages} {241101} (\bibinfo {year}
  {2011})},\ \Eprint {http://arxiv.org/abs/0909.2867} {arXiv:0909.2867 [gr-qc]}
  \BibitemShut {NoStop}%
\bibitem [{\citenamefont {Santamaria}\ \emph {et~al.}(2010)\citenamefont
  {Santamaria} \emph {et~al.}}]{Santamaria:2010yb}%
  \BibitemOpen
  \bibfield  {author} {\bibinfo {author} {\bibfnamefont {L.}~\bibnamefont
  {Santamaria}} \emph {et~al.},\ }\href {\doibase 10.1103/PhysRevD.82.064016}
  {\bibfield  {journal} {\bibinfo  {journal} {Phys. Rev. D}\ }\textbf {\bibinfo
  {volume} {82}},\ \bibinfo {pages} {064016} (\bibinfo {year} {2010})},\
  \Eprint {http://arxiv.org/abs/1005.3306} {arXiv:1005.3306 [gr-qc]}
  \BibitemShut {NoStop}%
\bibitem [{\citenamefont {Rodriguez}\ \emph
  {et~al.}(2016{\natexlab{b}})\citenamefont {Rodriguez}, \citenamefont {Zevin},
  \citenamefont {Pankow}, \citenamefont {Kalogera},\ and\ \citenamefont
  {Rasio}}]{Rodriguez2016}%
  \BibitemOpen
  \bibfield  {author} {\bibinfo {author} {\bibfnamefont {C.~L.}\ \bibnamefont
  {Rodriguez}}, \bibinfo {author} {\bibfnamefont {M.}~\bibnamefont {Zevin}},
  \bibinfo {author} {\bibfnamefont {C.}~\bibnamefont {Pankow}}, \bibinfo
  {author} {\bibfnamefont {V.}~\bibnamefont {Kalogera}}, \ and\ \bibinfo
  {author} {\bibfnamefont {F.~A.}\ \bibnamefont {Rasio}},\ }\href {\doibase
  10.3847/2041-8205/832/1/l2} {\bibfield  {journal} {\bibinfo  {journal} {The
  Astrophysical Journal}\ }\textbf {\bibinfo {volume} {832}},\ \bibinfo {pages}
  {L2} (\bibinfo {year} {2016}{\natexlab{b}})}\BibitemShut {NoStop}%
\bibitem [{\citenamefont {Gerosa}\ \emph {et~al.}(2018)\citenamefont {Gerosa},
  \citenamefont {Berti}, \citenamefont {O'Shaughnessy}, \citenamefont
  {Belczynski}, \citenamefont {Kesden}, \citenamefont {Wysocki},\ and\
  \citenamefont {Gladysz}}]{Gerosa2018}%
  \BibitemOpen
  \bibfield  {author} {\bibinfo {author} {\bibfnamefont {D.}~\bibnamefont
  {Gerosa}}, \bibinfo {author} {\bibfnamefont {E.}~\bibnamefont {Berti}},
  \bibinfo {author} {\bibfnamefont {R.}~\bibnamefont {O'Shaughnessy}}, \bibinfo
  {author} {\bibfnamefont {K.}~\bibnamefont {Belczynski}}, \bibinfo {author}
  {\bibfnamefont {M.}~\bibnamefont {Kesden}}, \bibinfo {author} {\bibfnamefont
  {D.}~\bibnamefont {Wysocki}}, \ and\ \bibinfo {author} {\bibfnamefont
  {W.}~\bibnamefont {Gladysz}},\ }\href {\doibase 10.1103/PhysRevD.98.084036}
  {\bibfield  {journal} {\bibinfo  {journal} {Phys. Rev. D}\ }\textbf {\bibinfo
  {volume} {98}},\ \bibinfo {pages} {084036} (\bibinfo {year}
  {2018})}\BibitemShut {NoStop}%
\bibitem [{\citenamefont {Belczynski}\ and\ \citenamefont
  {Bulik}(1999)}]{belczynski1999}%
  \BibitemOpen
  \bibfield  {author} {\bibinfo {author} {\bibfnamefont {K.}~\bibnamefont
  {Belczynski}}\ and\ \bibinfo {author} {\bibfnamefont {T.}~\bibnamefont
  {Bulik}},\ }\href@noop {} {\bibfield  {journal} {\bibinfo  {journal}
  {Astronomy and Astrophysics}\ }\textbf {\bibinfo {volume} {346}},\ \bibinfo
  {pages} {91} (\bibinfo {year} {1999})}\BibitemShut {NoStop}%
\bibitem [{\citenamefont {Callister}\ \emph {et~al.}(2020)\citenamefont
  {Callister}, \citenamefont {Farr},\ and\ \citenamefont
  {Renzo}}]{Callister2020}%
  \BibitemOpen
  \bibfield  {author} {\bibinfo {author} {\bibfnamefont {T.~A.}\ \bibnamefont
  {Callister}}, \bibinfo {author} {\bibfnamefont {W.~M.}\ \bibnamefont {Farr}},
  \ and\ \bibinfo {author} {\bibfnamefont {M.}~\bibnamefont {Renzo}},\
  }\href@noop {} {\enquote {\bibinfo {title} {State of the field: Binary black
  hole natal kicks and prospects for isolated field formation after
  {GWTC}-2},}\ } (\bibinfo {year} {2020}),\ \Eprint
  {http://arxiv.org/abs/arXiv:2011.09570} {arXiv:2011.09570} \BibitemShut
  {NoStop}%
\bibitem [{\citenamefont {Fuller}\ and\ \citenamefont {Ma}(2019)}]{Fuller2019}%
  \BibitemOpen
  \bibfield  {author} {\bibinfo {author} {\bibfnamefont {J.}~\bibnamefont
  {Fuller}}\ and\ \bibinfo {author} {\bibfnamefont {L.}~\bibnamefont {Ma}},\
  }\href {\doibase 10.3847/2041-8213/ab339b} {\bibfield  {journal} {\bibinfo
  {journal} {The Astrophysical Journal}\ }\textbf {\bibinfo {volume} {881}},\
  \bibinfo {pages} {L1} (\bibinfo {year} {2019})}\BibitemShut {NoStop}%
\bibitem [{\citenamefont {Bavera}\ \emph {et~al.}(2020)\citenamefont {Bavera},
  \citenamefont {Fragos}, \citenamefont {Qin}, \citenamefont {Zapartas},
  \citenamefont {Neijssel}, \citenamefont {Mandel}, \citenamefont {Batta},
  \citenamefont {Gaebel}, \citenamefont {Kimball},\ and\ \citenamefont
  {Stevenson}}]{Bavera2020}%
  \BibitemOpen
  \bibfield  {author} {\bibinfo {author} {\bibfnamefont {S.~S.}\ \bibnamefont
  {Bavera}}, \bibinfo {author} {\bibfnamefont {T.}~\bibnamefont {Fragos}},
  \bibinfo {author} {\bibfnamefont {Y.}~\bibnamefont {Qin}}, \bibinfo {author}
  {\bibfnamefont {E.}~\bibnamefont {Zapartas}}, \bibinfo {author}
  {\bibfnamefont {C.~J.}\ \bibnamefont {Neijssel}}, \bibinfo {author}
  {\bibfnamefont {I.}~\bibnamefont {Mandel}}, \bibinfo {author} {\bibfnamefont
  {A.}~\bibnamefont {Batta}}, \bibinfo {author} {\bibfnamefont {S.~M.}\
  \bibnamefont {Gaebel}}, \bibinfo {author} {\bibfnamefont {C.}~\bibnamefont
  {Kimball}}, \ and\ \bibinfo {author} {\bibfnamefont {S.}~\bibnamefont
  {Stevenson}},\ }\href {\doibase 10.1051/0004-6361/201936204} {\bibfield
  {journal} {\bibinfo  {journal} {Astronomy {\&} Astrophysics}\ }\textbf
  {\bibinfo {volume} {635}},\ \bibinfo {pages} {A97} (\bibinfo {year}
  {2020})}\BibitemShut {NoStop}%
\bibitem [{\citenamefont {Farr}\ \emph {et~al.}(2017)\citenamefont {Farr},
  \citenamefont {Stevenson}, \citenamefont {Miller}, \citenamefont {Mandel},
  \citenamefont {Farr},\ and\ \citenamefont {Vecchio}}]{Farr2017}%
  \BibitemOpen
  \bibfield  {author} {\bibinfo {author} {\bibfnamefont {W.~M.}\ \bibnamefont
  {Farr}}, \bibinfo {author} {\bibfnamefont {S.}~\bibnamefont {Stevenson}},
  \bibinfo {author} {\bibfnamefont {M.~C.}\ \bibnamefont {Miller}}, \bibinfo
  {author} {\bibfnamefont {I.}~\bibnamefont {Mandel}}, \bibinfo {author}
  {\bibfnamefont {B.}~\bibnamefont {Farr}}, \ and\ \bibinfo {author}
  {\bibfnamefont {A.}~\bibnamefont {Vecchio}},\ }\href {\doibase
  10.1038/nature23453} {\bibfield  {journal} {\bibinfo  {journal} {Nature}\
  }\textbf {\bibinfo {volume} {548}},\ \bibinfo {pages} {426} (\bibinfo {year}
  {2017})},\ \Eprint {http://arxiv.org/abs/1706.01385} {arXiv:1706.01385}
  \BibitemShut {NoStop}%
\bibitem [{\citenamefont {Farr}\ \emph {et~al.}(2018)\citenamefont {Farr},
  \citenamefont {Holz},\ and\ \citenamefont {Farr}}]{Farr2018}%
  \BibitemOpen
  \bibfield  {author} {\bibinfo {author} {\bibfnamefont {B.}~\bibnamefont
  {Farr}}, \bibinfo {author} {\bibfnamefont {D.~E.}\ \bibnamefont {Holz}}, \
  and\ \bibinfo {author} {\bibfnamefont {W.~M.}\ \bibnamefont {Farr}},\ }\href
  {\doibase 10.3847/2041-8213/aaaa64} {\bibfield  {journal} {\bibinfo
  {journal} {The Astrophysical Journal}\ }\textbf {\bibinfo {volume} {854}},\
  \bibinfo {pages} {L9} (\bibinfo {year} {2018})}\BibitemShut {NoStop}%
\bibitem [{\citenamefont {Roulet}\ and\ \citenamefont
  {Zaldarriaga}(2019)}]{Roulet2019}%
  \BibitemOpen
  \bibfield  {author} {\bibinfo {author} {\bibfnamefont {J.}~\bibnamefont
  {Roulet}}\ and\ \bibinfo {author} {\bibfnamefont {M.}~\bibnamefont
  {Zaldarriaga}},\ }\href {\doibase 10.1093/mnras/stz226} {\bibfield  {journal}
  {\bibinfo  {journal} {Monthly Notices of the Royal Astronomical Society}\
  }\textbf {\bibinfo {volume} {484}},\ \bibinfo {pages} {4216} (\bibinfo {year}
  {2019})}\BibitemShut {NoStop}%
\bibitem [{\citenamefont {Roulet}\ \emph {et~al.}(2020)\citenamefont {Roulet},
  \citenamefont {Venumadhav}, \citenamefont {Zackay}, \citenamefont {Dai},\
  and\ \citenamefont {Zaldarriaga}}]{Roulet2020}%
  \BibitemOpen
  \bibfield  {author} {\bibinfo {author} {\bibfnamefont {J.}~\bibnamefont
  {Roulet}}, \bibinfo {author} {\bibfnamefont {T.}~\bibnamefont {Venumadhav}},
  \bibinfo {author} {\bibfnamefont {B.}~\bibnamefont {Zackay}}, \bibinfo
  {author} {\bibfnamefont {L.}~\bibnamefont {Dai}}, \ and\ \bibinfo {author}
  {\bibfnamefont {M.}~\bibnamefont {Zaldarriaga}},\ }\href {\doibase
  10.1103/PhysRevD.102.123022} {\bibfield  {journal} {\bibinfo  {journal}
  {Phys. Rev. D}\ }\textbf {\bibinfo {volume} {102}},\ \bibinfo {pages}
  {123022} (\bibinfo {year} {2020})}\BibitemShut {NoStop}%
\bibitem [{\citenamefont {Fowler}\ and\ \citenamefont
  {Hoyle}(1964)}]{Fowler1964}%
  \BibitemOpen
  \bibfield  {author} {\bibinfo {author} {\bibfnamefont {W.~A.}\ \bibnamefont
  {Fowler}}\ and\ \bibinfo {author} {\bibfnamefont {F.}~\bibnamefont {Hoyle}},\
  }\href {\doibase 10.1086/190103} {\bibfield  {journal} {\bibinfo  {journal}
  {The Astrophysical Journal Supplement Series}\ }\textbf {\bibinfo {volume}
  {9}},\ \bibinfo {pages} {201} (\bibinfo {year} {1964})}\BibitemShut {NoStop}%
\bibitem [{\citenamefont {Barkat}\ \emph {et~al.}(1967)\citenamefont {Barkat},
  \citenamefont {Rakavy},\ and\ \citenamefont {Sack}}]{Barkat1967}%
  \BibitemOpen
  \bibfield  {author} {\bibinfo {author} {\bibfnamefont {Z.}~\bibnamefont
  {Barkat}}, \bibinfo {author} {\bibfnamefont {G.}~\bibnamefont {Rakavy}}, \
  and\ \bibinfo {author} {\bibfnamefont {N.}~\bibnamefont {Sack}},\ }\href
  {\doibase 10.1103/PhysRevLett.18.379} {\bibfield  {journal} {\bibinfo
  {journal} {Phys. Rev. Lett.}\ }\textbf {\bibinfo {volume} {18}},\ \bibinfo
  {pages} {379} (\bibinfo {year} {1967})}\BibitemShut {NoStop}%
\bibitem [{\citenamefont {Bond}\ \emph {et~al.}(1984)\citenamefont {Bond},
  \citenamefont {Arnett},\ and\ \citenamefont {Carr}}]{Bond1984}%
  \BibitemOpen
  \bibfield  {author} {\bibinfo {author} {\bibfnamefont {J.~R.}\ \bibnamefont
  {Bond}}, \bibinfo {author} {\bibfnamefont {W.~D.}\ \bibnamefont {Arnett}}, \
  and\ \bibinfo {author} {\bibfnamefont {B.~J.}\ \bibnamefont {Carr}},\ }\href
  {\doibase 10.1086/162057} {\bibfield  {journal} {\bibinfo  {journal} {The
  Astrophysical Journal}\ }\textbf {\bibinfo {volume} {280}},\ \bibinfo {pages}
  {825} (\bibinfo {year} {1984})}\BibitemShut {NoStop}%
\bibitem [{\citenamefont {Heger}\ \emph {et~al.}(2003)\citenamefont {Heger},
  \citenamefont {Fryer}, \citenamefont {Woosley}, \citenamefont {Langer},\ and\
  \citenamefont {Hartmann}}]{Heger2003}%
  \BibitemOpen
  \bibfield  {author} {\bibinfo {author} {\bibfnamefont {A.}~\bibnamefont
  {Heger}}, \bibinfo {author} {\bibfnamefont {C.~L.}\ \bibnamefont {Fryer}},
  \bibinfo {author} {\bibfnamefont {S.~E.}\ \bibnamefont {Woosley}}, \bibinfo
  {author} {\bibfnamefont {N.}~\bibnamefont {Langer}}, \ and\ \bibinfo {author}
  {\bibfnamefont {D.~H.}\ \bibnamefont {Hartmann}},\ }\href {\doibase
  10.1086/375341} {\bibfield  {journal} {\bibinfo  {journal} {The Astrophysical
  Journal}\ }\textbf {\bibinfo {volume} {591}},\ \bibinfo {pages} {288}
  (\bibinfo {year} {2003})}\BibitemShut {NoStop}%
\bibitem [{\citenamefont {Farmer}\ \emph {et~al.}(2019)\citenamefont {Farmer},
  \citenamefont {Renzo}, \citenamefont {de~Mink}, \citenamefont {Marchant},\
  and\ \citenamefont {Justham}}]{Farmer2019}%
  \BibitemOpen
  \bibfield  {author} {\bibinfo {author} {\bibfnamefont {R.}~\bibnamefont
  {Farmer}}, \bibinfo {author} {\bibfnamefont {M.}~\bibnamefont {Renzo}},
  \bibinfo {author} {\bibfnamefont {S.~E.}\ \bibnamefont {de~Mink}}, \bibinfo
  {author} {\bibfnamefont {P.}~\bibnamefont {Marchant}}, \ and\ \bibinfo
  {author} {\bibfnamefont {S.}~\bibnamefont {Justham}},\ }\href {\doibase
  10.3847/1538-4357/ab518b} {\bibfield  {journal} {\bibinfo  {journal} {The
  Astrophysical Journal}\ }\textbf {\bibinfo {volume} {887}},\ \bibinfo {pages}
  {53} (\bibinfo {year} {2019})}\BibitemShut {NoStop}%
\bibitem [{\citenamefont {Safarzadeh}\ and\ \citenamefont
  {Haiman}(2020)}]{Safarzadeh2020}%
  \BibitemOpen
  \bibfield  {author} {\bibinfo {author} {\bibfnamefont {M.}~\bibnamefont
  {Safarzadeh}}\ and\ \bibinfo {author} {\bibfnamefont {Z.}~\bibnamefont
  {Haiman}},\ }\href {\doibase 10.3847/2041-8213/abc253} {\bibfield  {journal}
  {\bibinfo  {journal} {The Astrophysical Journal}\ }\textbf {\bibinfo {volume}
  {903}},\ \bibinfo {pages} {L21} (\bibinfo {year} {2020})}\BibitemShut
  {NoStop}%
\bibitem [{\citenamefont {Gerosa}\ and\ \citenamefont
  {Fishbach}(2021)}]{Gerosa2021}%
  \BibitemOpen
  \bibfield  {author} {\bibinfo {author} {\bibfnamefont {D.}~\bibnamefont
  {Gerosa}}\ and\ \bibinfo {author} {\bibfnamefont {M.}~\bibnamefont
  {Fishbach}},\ }\href@noop {} {\enquote {\bibinfo {title} {Hierarchical
  mergers of stellar-mass black holes and their gravitational-wave
  signatures},}\ } (\bibinfo {year} {2021}),\ \Eprint
  {http://arxiv.org/abs/arXiv:2105.03439} {arXiv:2105.03439} \BibitemShut
  {NoStop}%
\bibitem [{\citenamefont {Fishbach}\ and\ \citenamefont
  {Holz}(2017)}]{Fishbach2017}%
  \BibitemOpen
  \bibfield  {author} {\bibinfo {author} {\bibfnamefont {M.}~\bibnamefont
  {Fishbach}}\ and\ \bibinfo {author} {\bibfnamefont {D.~E.}\ \bibnamefont
  {Holz}},\ }\href {\doibase 10.3847/2041-8213/aa9bf6} {\bibfield  {journal}
  {\bibinfo  {journal} {The Astrophysical Journal}\ }\textbf {\bibinfo {volume}
  {851}},\ \bibinfo {pages} {L25} (\bibinfo {year} {2017})}\BibitemShut
  {NoStop}%
\bibitem [{\citenamefont {Wysocki}\ \emph {et~al.}(2019)\citenamefont
  {Wysocki}, \citenamefont {Lange},\ and\ \citenamefont
  {O'Shaughnessy}}]{Wysocki2019}%
  \BibitemOpen
  \bibfield  {author} {\bibinfo {author} {\bibfnamefont {D.}~\bibnamefont
  {Wysocki}}, \bibinfo {author} {\bibfnamefont {J.}~\bibnamefont {Lange}}, \
  and\ \bibinfo {author} {\bibfnamefont {R.}~\bibnamefont {O'Shaughnessy}},\
  }\href {\doibase 10.1103/PhysRevD.100.043012} {\bibfield  {journal} {\bibinfo
   {journal} {Phys. Rev. D}\ }\textbf {\bibinfo {volume} {100}},\ \bibinfo
  {pages} {043012} (\bibinfo {year} {2019})}\BibitemShut {NoStop}%
\bibitem [{\citenamefont {Abbott}\ \emph
  {et~al.}(2020{\natexlab{c}})\citenamefont {Abbott} \emph
  {et~al.}}]{Abbott:2020khf}%
  \BibitemOpen
  \bibfield  {author} {\bibinfo {author} {\bibfnamefont {R.}~\bibnamefont
  {Abbott}} \emph {et~al.} (\bibinfo {collaboration} {LIGO Scientific,
  Virgo}),\ }\href {\doibase 10.3847/2041-8213/ab960f} {\bibfield  {journal}
  {\bibinfo  {journal} {Astrophys. J. Lett.}\ }\textbf {\bibinfo {volume}
  {896}},\ \bibinfo {pages} {L44} (\bibinfo {year} {2020}{\natexlab{c}})},\
  \Eprint {http://arxiv.org/abs/2006.12611} {arXiv:2006.12611 [astro-ph.HE]}
  \BibitemShut {NoStop}%
\bibitem [{\citenamefont {Nitz}\ \emph {et~al.}(2021)\citenamefont {Nitz},
  \citenamefont {Capano}, \citenamefont {Kumar}, \citenamefont {Wang},
  \citenamefont {Kastha}, \citenamefont {Schäfer}, \citenamefont {Dhurkunde},\
  and\ \citenamefont {Cabero}}]{3OGC}%
  \BibitemOpen
  \bibfield  {author} {\bibinfo {author} {\bibfnamefont {A.~H.}\ \bibnamefont
  {Nitz}}, \bibinfo {author} {\bibfnamefont {C.~D.}\ \bibnamefont {Capano}},
  \bibinfo {author} {\bibfnamefont {S.}~\bibnamefont {Kumar}}, \bibinfo
  {author} {\bibfnamefont {Y.-F.}\ \bibnamefont {Wang}}, \bibinfo {author}
  {\bibfnamefont {S.}~\bibnamefont {Kastha}}, \bibinfo {author} {\bibfnamefont
  {M.}~\bibnamefont {Schäfer}}, \bibinfo {author} {\bibfnamefont
  {R.}~\bibnamefont {Dhurkunde}}, \ and\ \bibinfo {author} {\bibfnamefont
  {M.}~\bibnamefont {Cabero}},\ }\href@noop {} {\enquote {\bibinfo {title}
  {3-{OGC}: Catalog of gravitational waves from compact-binary mergers},}\ }
  (\bibinfo {year} {2021}),\ \Eprint {http://arxiv.org/abs/arXiv:2105.09151}
  {arXiv:2105.09151} \BibitemShut {NoStop}%
\bibitem [{\citenamefont {Pratten}\ \emph {et~al.}(2020)\citenamefont
  {Pratten}, \citenamefont {García-Quirós}, \citenamefont {Colleoni},
  \citenamefont {Ramos-Buades}, \citenamefont {Estellés}, \citenamefont
  {Mateu-Lucena}, \citenamefont {Jaume}, \citenamefont {Haney}, \citenamefont
  {Keitel}, \citenamefont {Thompson},\ and\ \citenamefont
  {Husa}}]{Pratten2020}%
  \BibitemOpen
  \bibfield  {author} {\bibinfo {author} {\bibfnamefont {G.}~\bibnamefont
  {Pratten}}, \bibinfo {author} {\bibfnamefont {C.}~\bibnamefont
  {García-Quirós}}, \bibinfo {author} {\bibfnamefont {M.}~\bibnamefont
  {Colleoni}}, \bibinfo {author} {\bibfnamefont {A.}~\bibnamefont
  {Ramos-Buades}}, \bibinfo {author} {\bibfnamefont {H.}~\bibnamefont
  {Estellés}}, \bibinfo {author} {\bibfnamefont {M.}~\bibnamefont
  {Mateu-Lucena}}, \bibinfo {author} {\bibfnamefont {R.}~\bibnamefont {Jaume}},
  \bibinfo {author} {\bibfnamefont {M.}~\bibnamefont {Haney}}, \bibinfo
  {author} {\bibfnamefont {D.}~\bibnamefont {Keitel}}, \bibinfo {author}
  {\bibfnamefont {J.~E.}\ \bibnamefont {Thompson}}, \ and\ \bibinfo {author}
  {\bibfnamefont {S.}~\bibnamefont {Husa}},\ }\href@noop {} {\enquote {\bibinfo
  {title} {Computationally efficient models for the dominant and sub-dominant
  harmonic modes of precessing binary black holes},}\ } (\bibinfo {year}
  {2020}),\ \Eprint {http://arxiv.org/abs/arXiv:2004.06503} {arXiv:2004.06503}
  \BibitemShut {NoStop}%
\bibitem [{\citenamefont {Zackay}\ \emph {et~al.}(2018)\citenamefont {Zackay},
  \citenamefont {Dai},\ and\ \citenamefont {Venumadhav}}]{Zackay2018}%
  \BibitemOpen
  \bibfield  {author} {\bibinfo {author} {\bibfnamefont {B.}~\bibnamefont
  {Zackay}}, \bibinfo {author} {\bibfnamefont {L.}~\bibnamefont {Dai}}, \ and\
  \bibinfo {author} {\bibfnamefont {T.}~\bibnamefont {Venumadhav}},\
  }\href@noop {} {\enquote {\bibinfo {title} {Relative binning and fast
  likelihood evaluation for gravitational wave parameter estimation},}\ }
  (\bibinfo {year} {2018}),\ \Eprint {http://arxiv.org/abs/1806.08792}
  {arXiv:1806.08792 [astro-ph.IM]} \BibitemShut {NoStop}%
\bibitem [{\citenamefont {Buchner}\ \emph {et~al.}(2014)\citenamefont
  {Buchner}, \citenamefont {Georgakakis}, \citenamefont {Nandra}, \citenamefont
  {Hsu}, \citenamefont {Rangel}, \citenamefont {Brightman}, \citenamefont
  {Merloni}, \citenamefont {Salvato}, \citenamefont {Donley},\ and\
  \citenamefont {Kocevski}}]{Buchner2014}%
  \BibitemOpen
  \bibfield  {author} {\bibinfo {author} {\bibfnamefont {J.}~\bibnamefont
  {Buchner}}, \bibinfo {author} {\bibfnamefont {A.}~\bibnamefont
  {Georgakakis}}, \bibinfo {author} {\bibfnamefont {K.}~\bibnamefont {Nandra}},
  \bibinfo {author} {\bibfnamefont {L.}~\bibnamefont {Hsu}}, \bibinfo {author}
  {\bibfnamefont {C.}~\bibnamefont {Rangel}}, \bibinfo {author} {\bibfnamefont
  {M.}~\bibnamefont {Brightman}}, \bibinfo {author} {\bibfnamefont
  {A.}~\bibnamefont {Merloni}}, \bibinfo {author} {\bibfnamefont
  {M.}~\bibnamefont {Salvato}}, \bibinfo {author} {\bibfnamefont
  {J.}~\bibnamefont {Donley}}, \ and\ \bibinfo {author} {\bibfnamefont
  {D.}~\bibnamefont {Kocevski}},\ }\href {\doibase 10.1051/0004-6361/201322971}
  {\bibfield  {journal} {\bibinfo  {journal} {Astronomy {\&} Astrophysics}\
  }\textbf {\bibinfo {volume} {564}},\ \bibinfo {pages} {A125} (\bibinfo {year}
  {2014})}\BibitemShut {NoStop}%
\bibitem [{\citenamefont {Cornish}\ and\ \citenamefont
  {Littenberg}(2015)}]{Cornish2015}%
  \BibitemOpen
  \bibfield  {author} {\bibinfo {author} {\bibfnamefont {N.~J.}\ \bibnamefont
  {Cornish}}\ and\ \bibinfo {author} {\bibfnamefont {T.~B.}\ \bibnamefont
  {Littenberg}},\ }\href {\doibase 10.1088/0264-9381/32/13/135012} {\bibfield
  {journal} {\bibinfo  {journal} {Classical and Quantum Gravity}\ }\textbf
  {\bibinfo {volume} {32}},\ \bibinfo {pages} {135012} (\bibinfo {year}
  {2015})}\BibitemShut {NoStop}%
\bibitem [{\citenamefont {Littenberg}\ \emph {et~al.}(2016)\citenamefont
  {Littenberg}, \citenamefont {Kanner}, \citenamefont {Cornish},\ and\
  \citenamefont {Millhouse}}]{Littenberg2016}%
  \BibitemOpen
  \bibfield  {author} {\bibinfo {author} {\bibfnamefont {T.~B.}\ \bibnamefont
  {Littenberg}}, \bibinfo {author} {\bibfnamefont {J.~B.}\ \bibnamefont
  {Kanner}}, \bibinfo {author} {\bibfnamefont {N.~J.}\ \bibnamefont {Cornish}},
  \ and\ \bibinfo {author} {\bibfnamefont {M.}~\bibnamefont {Millhouse}},\
  }\href {\doibase 10.1103/PhysRevD.94.044050} {\bibfield  {journal} {\bibinfo
  {journal} {Phys. Rev. D}\ }\textbf {\bibinfo {volume} {94}},\ \bibinfo
  {pages} {044050} (\bibinfo {year} {2016})}\BibitemShut {NoStop}%
\bibitem [{\citenamefont {Cornish}\ \emph {et~al.}(2021)\citenamefont
  {Cornish}, \citenamefont {Littenberg}, \citenamefont {B\'ecsy}, \citenamefont
  {Chatziioannou}, \citenamefont {Clark}, \citenamefont {Ghonge},\ and\
  \citenamefont {Millhouse}}]{Cornish2021}%
  \BibitemOpen
  \bibfield  {author} {\bibinfo {author} {\bibfnamefont {N.~J.}\ \bibnamefont
  {Cornish}}, \bibinfo {author} {\bibfnamefont {T.~B.}\ \bibnamefont
  {Littenberg}}, \bibinfo {author} {\bibfnamefont {B.}~\bibnamefont {B\'ecsy}},
  \bibinfo {author} {\bibfnamefont {K.}~\bibnamefont {Chatziioannou}}, \bibinfo
  {author} {\bibfnamefont {J.~A.}\ \bibnamefont {Clark}}, \bibinfo {author}
  {\bibfnamefont {S.}~\bibnamefont {Ghonge}}, \ and\ \bibinfo {author}
  {\bibfnamefont {M.}~\bibnamefont {Millhouse}},\ }\href {\doibase
  10.1103/PhysRevD.103.044006} {\bibfield  {journal} {\bibinfo  {journal}
  {Phys. Rev. D}\ }\textbf {\bibinfo {volume} {103}},\ \bibinfo {pages}
  {044006} (\bibinfo {year} {2021})}\BibitemShut {NoStop}%
\bibitem [{\citenamefont {Abbott}\ \emph {et~al.}(2016)\citenamefont {Abbott}
  \emph {et~al.}}]{Abbott:2016nmj}%
  \BibitemOpen
  \bibfield  {author} {\bibinfo {author} {\bibfnamefont {B.~P.}\ \bibnamefont
  {Abbott}} \emph {et~al.} (\bibinfo {collaboration} {LIGO Scientific,
  Virgo}),\ }\href {\doibase 10.1103/PhysRevLett.116.241103} {\bibfield
  {journal} {\bibinfo  {journal} {Phys. Rev. Lett.}\ }\textbf {\bibinfo
  {volume} {116}},\ \bibinfo {pages} {241103} (\bibinfo {year} {2016})},\
  \Eprint {http://arxiv.org/abs/1606.04855} {arXiv:1606.04855 [gr-qc]}
  \BibitemShut {NoStop}%
\bibitem [{\citenamefont {Chia}\ \emph {et~al.}(2021)\citenamefont {Chia},
  \citenamefont {Olsen}, \citenamefont {Roulet}, \citenamefont {Dai},
  \citenamefont {Venumadhav}, \citenamefont {Zackay},\ and\ \citenamefont
  {Zaldarriaga}}]{Chia2021}%
  \BibitemOpen
  \bibfield  {author} {\bibinfo {author} {\bibfnamefont {H.~S.}\ \bibnamefont
  {Chia}}, \bibinfo {author} {\bibfnamefont {S.}~\bibnamefont {Olsen}},
  \bibinfo {author} {\bibfnamefont {J.}~\bibnamefont {Roulet}}, \bibinfo
  {author} {\bibfnamefont {L.}~\bibnamefont {Dai}}, \bibinfo {author}
  {\bibfnamefont {T.}~\bibnamefont {Venumadhav}}, \bibinfo {author}
  {\bibfnamefont {B.}~\bibnamefont {Zackay}}, \ and\ \bibinfo {author}
  {\bibfnamefont {M.}~\bibnamefont {Zaldarriaga}},\ }\href@noop {} {\enquote
  {\bibinfo {title} {Boxing day surprise: Higher multipoles and orbital
  precession in {GW}151226},}\ } (\bibinfo {year} {2021}),\ \Eprint
  {http://arxiv.org/abs/arXiv:2105.06486} {arXiv:2105.06486} \BibitemShut
  {NoStop}%
\bibitem [{\citenamefont {Abbott}\ \emph
  {et~al.}(2020{\natexlab{d}})\citenamefont {Abbott} \emph
  {et~al.}}]{GW190521}%
  \BibitemOpen
  \bibfield  {author} {\bibinfo {author} {\bibfnamefont {R.}~\bibnamefont
  {Abbott}} \emph {et~al.},\ }\href {\doibase 10.1103/physrevlett.125.101102}
  {\bibfield  {journal} {\bibinfo  {journal} {Physical Review Letters}\
  }\textbf {\bibinfo {volume} {125}} (\bibinfo {year} {2020}{\natexlab{d}}),\
  10.1103/physrevlett.125.101102}\BibitemShut {NoStop}%
\bibitem [{\citenamefont {Nitz}\ and\ \citenamefont {Capano}(2021)}]{Nitz2021}%
  \BibitemOpen
  \bibfield  {author} {\bibinfo {author} {\bibfnamefont {A.~H.}\ \bibnamefont
  {Nitz}}\ and\ \bibinfo {author} {\bibfnamefont {C.~D.}\ \bibnamefont
  {Capano}},\ }\href {\doibase 10.3847/2041-8213/abccc5} {\bibfield  {journal}
  {\bibinfo  {journal} {The Astrophysical Journal}\ }\textbf {\bibinfo {volume}
  {907}},\ \bibinfo {pages} {L9} (\bibinfo {year} {2021})}\BibitemShut
  {NoStop}%
\bibitem [{\citenamefont {Olsen}\ \emph {et~al.}(2021)\citenamefont {Olsen},
  \citenamefont {Roulet}, \citenamefont {Chia}, \citenamefont {Dai},
  \citenamefont {Venumadhav}, \citenamefont {Zackay},\ and\ \citenamefont
  {Zaldarriaga}}]{Olsen2021}%
  \BibitemOpen
  \bibfield  {author} {\bibinfo {author} {\bibfnamefont {S.}~\bibnamefont
  {Olsen}}, \bibinfo {author} {\bibfnamefont {J.}~\bibnamefont {Roulet}},
  \bibinfo {author} {\bibfnamefont {H.~S.}\ \bibnamefont {Chia}}, \bibinfo
  {author} {\bibfnamefont {L.}~\bibnamefont {Dai}}, \bibinfo {author}
  {\bibfnamefont {T.}~\bibnamefont {Venumadhav}}, \bibinfo {author}
  {\bibfnamefont {B.}~\bibnamefont {Zackay}}, \ and\ \bibinfo {author}
  {\bibfnamefont {M.}~\bibnamefont {Zaldarriaga}},\ }\href@noop {} {\enquote
  {\bibinfo {title} {Mapping the likelihood of {GW}190521 with diverse mass and
  spin priors},}\ } (\bibinfo {year} {2021}),\ \Eprint
  {http://arxiv.org/abs/2106.13821} {arXiv:2106.13821 [astro-ph.HE]}
  \BibitemShut {NoStop}%
\bibitem [{\citenamefont {Estellés}\ \emph {et~al.}(2021)\citenamefont
  {Estellés}, \citenamefont {Husa}, \citenamefont {Colleoni}, \citenamefont
  {Mateu-Lucena}, \citenamefont {de~Lluc~Planas}, \citenamefont
  {García-Quirós}, \citenamefont {Keitel}, \citenamefont {Ramos-Buades},
  \citenamefont {Mehta}, \citenamefont {Buonanno},\ and\ \citenamefont
  {Ossokine}}]{Estelles2021}%
  \BibitemOpen
  \bibfield  {author} {\bibinfo {author} {\bibfnamefont {H.}~\bibnamefont
  {Estellés}}, \bibinfo {author} {\bibfnamefont {S.}~\bibnamefont {Husa}},
  \bibinfo {author} {\bibfnamefont {M.}~\bibnamefont {Colleoni}}, \bibinfo
  {author} {\bibfnamefont {M.}~\bibnamefont {Mateu-Lucena}}, \bibinfo {author}
  {\bibfnamefont {M.}~\bibnamefont {de~Lluc~Planas}}, \bibinfo {author}
  {\bibfnamefont {C.}~\bibnamefont {García-Quirós}}, \bibinfo {author}
  {\bibfnamefont {D.}~\bibnamefont {Keitel}}, \bibinfo {author} {\bibfnamefont
  {A.}~\bibnamefont {Ramos-Buades}}, \bibinfo {author} {\bibfnamefont {A.~K.}\
  \bibnamefont {Mehta}}, \bibinfo {author} {\bibfnamefont {A.}~\bibnamefont
  {Buonanno}}, \ and\ \bibinfo {author} {\bibfnamefont {S.}~\bibnamefont
  {Ossokine}},\ }\href@noop {} {\enquote {\bibinfo {title} {A detailed analysis
  of {GW}190521 with phenomenological waveform models},}\ } (\bibinfo {year}
  {2021}),\ \Eprint {http://arxiv.org/abs/arXiv:2105.06360} {arXiv:2105.06360}
  \BibitemShut {NoStop}%
\bibitem [{\citenamefont {Campanelli}\ \emph {et~al.}(2006)\citenamefont
  {Campanelli}, \citenamefont {Lousto},\ and\ \citenamefont
  {Zlochower}}]{Campanelli2006}%
  \BibitemOpen
  \bibfield  {author} {\bibinfo {author} {\bibfnamefont {M.}~\bibnamefont
  {Campanelli}}, \bibinfo {author} {\bibfnamefont {C.~O.}\ \bibnamefont
  {Lousto}}, \ and\ \bibinfo {author} {\bibfnamefont {Y.}~\bibnamefont
  {Zlochower}},\ }\href {\doibase 10.1103/physrevd.74.041501} {\bibfield
  {journal} {\bibinfo  {journal} {Physical Review D}\ }\textbf {\bibinfo
  {volume} {74}} (\bibinfo {year} {2006}),\
  10.1103/physrevd.74.041501}\BibitemShut {NoStop}%
\bibitem [{\citenamefont {Ng}\ \emph {et~al.}(2018)\citenamefont {Ng},
  \citenamefont {Vitale}, \citenamefont {Zimmerman}, \citenamefont
  {Chatziioannou}, \citenamefont {Gerosa},\ and\ \citenamefont
  {Haster}}]{Ng2018}%
  \BibitemOpen
  \bibfield  {author} {\bibinfo {author} {\bibfnamefont {K.~K.~Y.}\
  \bibnamefont {Ng}}, \bibinfo {author} {\bibfnamefont {S.}~\bibnamefont
  {Vitale}}, \bibinfo {author} {\bibfnamefont {A.}~\bibnamefont {Zimmerman}},
  \bibinfo {author} {\bibfnamefont {K.}~\bibnamefont {Chatziioannou}}, \bibinfo
  {author} {\bibfnamefont {D.}~\bibnamefont {Gerosa}}, \ and\ \bibinfo {author}
  {\bibfnamefont {C.-J.}\ \bibnamefont {Haster}},\ }\href {\doibase
  10.1103/PhysRevD.98.083007} {\bibfield  {journal} {\bibinfo  {journal} {Phys.
  Rev. D}\ }\textbf {\bibinfo {volume} {98}},\ \bibinfo {pages} {083007}
  (\bibinfo {year} {2018})}\BibitemShut {NoStop}%
\bibitem [{\citenamefont {Kushnir}\ \emph {et~al.}(2016)\citenamefont
  {Kushnir}, \citenamefont {Zaldarriaga}, \citenamefont {Kollmeier},\ and\
  \citenamefont {Waldman}}]{Kushnir2016}%
  \BibitemOpen
  \bibfield  {author} {\bibinfo {author} {\bibfnamefont {D.}~\bibnamefont
  {Kushnir}}, \bibinfo {author} {\bibfnamefont {M.}~\bibnamefont
  {Zaldarriaga}}, \bibinfo {author} {\bibfnamefont {J.~A.}\ \bibnamefont
  {Kollmeier}}, \ and\ \bibinfo {author} {\bibfnamefont {R.}~\bibnamefont
  {Waldman}},\ }\href {\doibase 10.1093/mnras/stw1684} {\bibfield  {journal}
  {\bibinfo  {journal} {Monthly Notices of the Royal Astronomical Society}\
  }\textbf {\bibinfo {volume} {462}},\ \bibinfo {pages} {844} (\bibinfo {year}
  {2016})}\BibitemShut {NoStop}%
\bibitem [{\citenamefont {Zaldarriaga}\ \emph {et~al.}(2017)\citenamefont
  {Zaldarriaga}, \citenamefont {Kushnir},\ and\ \citenamefont
  {Kollmeier}}]{Zaldarriaga2017}%
  \BibitemOpen
  \bibfield  {author} {\bibinfo {author} {\bibfnamefont {M.}~\bibnamefont
  {Zaldarriaga}}, \bibinfo {author} {\bibfnamefont {D.}~\bibnamefont
  {Kushnir}}, \ and\ \bibinfo {author} {\bibfnamefont {J.~A.}\ \bibnamefont
  {Kollmeier}},\ }\href {\doibase 10.1093/mnras/stx2577} {\bibfield  {journal}
  {\bibinfo  {journal} {Monthly Notices of the Royal Astronomical Society}\
  }\textbf {\bibinfo {volume} {473}},\ \bibinfo {pages} {4174} (\bibinfo {year}
  {2017})}\BibitemShut {NoStop}%
\bibitem [{\citenamefont {Qin}\ \emph {et~al.}(2018)\citenamefont {Qin},
  \citenamefont {Fragos}, \citenamefont {Meynet}, \citenamefont {Andrews},
  \citenamefont {S{\o}rensen},\ and\ \citenamefont {Song}}]{Qin2018}%
  \BibitemOpen
  \bibfield  {author} {\bibinfo {author} {\bibfnamefont {Y.}~\bibnamefont
  {Qin}}, \bibinfo {author} {\bibfnamefont {T.}~\bibnamefont {Fragos}},
  \bibinfo {author} {\bibfnamefont {G.}~\bibnamefont {Meynet}}, \bibinfo
  {author} {\bibfnamefont {J.}~\bibnamefont {Andrews}}, \bibinfo {author}
  {\bibfnamefont {M.}~\bibnamefont {S{\o}rensen}}, \ and\ \bibinfo {author}
  {\bibfnamefont {H.~F.}\ \bibnamefont {Song}},\ }\href {\doibase
  10.1051/0004-6361/201832839} {\bibfield  {journal} {\bibinfo  {journal}
  {Astronomy {\&} Astrophysics}\ }\textbf {\bibinfo {volume} {616}},\ \bibinfo
  {pages} {A28} (\bibinfo {year} {2018})}\BibitemShut {NoStop}%
\bibitem [{\citenamefont {Abbott}\ \emph {et~al.}(2021)\citenamefont {Abbott}
  \emph {et~al.}}]{GWOSC}%
  \BibitemOpen
  \bibfield  {author} {\bibinfo {author} {\bibfnamefont {R.}~\bibnamefont
  {Abbott}} \emph {et~al.},\ }\href {\doibase 10.1016/j.softx.2021.100658}
  {\bibfield  {journal} {\bibinfo  {journal} {{SoftwareX}}\ }\textbf {\bibinfo
  {volume} {13}},\ \bibinfo {pages} {100658} (\bibinfo {year}
  {2021})}\BibitemShut {NoStop}%
\bibitem [{\citenamefont {Venumadhav}\ \emph
  {et~al.}(2019{\natexlab{b}})\citenamefont {Venumadhav}, \citenamefont
  {Zackay}, \citenamefont {Roulet}, \citenamefont {Dai},\ and\ \citenamefont
  {Zaldarriaga}}]{pipeline}%
  \BibitemOpen
  \bibfield  {author} {\bibinfo {author} {\bibfnamefont {T.}~\bibnamefont
  {Venumadhav}}, \bibinfo {author} {\bibfnamefont {B.}~\bibnamefont {Zackay}},
  \bibinfo {author} {\bibfnamefont {J.}~\bibnamefont {Roulet}}, \bibinfo
  {author} {\bibfnamefont {L.}~\bibnamefont {Dai}}, \ and\ \bibinfo {author}
  {\bibfnamefont {M.}~\bibnamefont {Zaldarriaga}},\ }\href {\doibase
  10.1103/PhysRevD.100.023011} {\bibfield  {journal} {\bibinfo  {journal}
  {Phys. Rev. D}\ }\textbf {\bibinfo {volume} {100}},\ \bibinfo {pages}
  {023011} (\bibinfo {year} {2019}{\natexlab{b}})}\BibitemShut {NoStop}%
\bibitem [{\citenamefont {Venumadhav}\ \emph
  {et~al.}(2020{\natexlab{b}})\citenamefont {Venumadhav}, \citenamefont
  {Zackay}, \citenamefont {Roulet}, \citenamefont {Dai},\ and\ \citenamefont
  {Zaldarriaga}}]{BBH_O2}%
  \BibitemOpen
  \bibfield  {author} {\bibinfo {author} {\bibfnamefont {T.}~\bibnamefont
  {Venumadhav}}, \bibinfo {author} {\bibfnamefont {B.}~\bibnamefont {Zackay}},
  \bibinfo {author} {\bibfnamefont {J.}~\bibnamefont {Roulet}}, \bibinfo
  {author} {\bibfnamefont {L.}~\bibnamefont {Dai}}, \ and\ \bibinfo {author}
  {\bibfnamefont {M.}~\bibnamefont {Zaldarriaga}},\ }\href {\doibase
  10.1103/PhysRevD.101.083030} {\bibfield  {journal} {\bibinfo  {journal}
  {Phys. Rev. D}\ }\textbf {\bibinfo {volume} {101}},\ \bibinfo {pages}
  {083030} (\bibinfo {year} {2020}{\natexlab{b}})}\BibitemShut {NoStop}%
\bibitem [{gwo()}]{gwosc_url}%
  \BibitemOpen
  \href@noop {} {\enquote {\bibinfo {title} {{Gravitational Wave Open Science
  Center (GWOSC)}},}\ }\bibinfo {howpublished}
  {\url{www.gw-openscience.org/O2/}}\BibitemShut {NoStop}%
\bibitem [{\citenamefont {Abbott}\ \emph
  {et~al.}(2019{\natexlab{b}})\citenamefont {Abbott} \emph
  {et~al.}}]{GWTC1_pop}%
  \BibitemOpen
  \bibfield  {author} {\bibinfo {author} {\bibfnamefont {B.~P.}\ \bibnamefont
  {Abbott}} \emph {et~al.},\ }\href {\doibase 10.3847/2041-8213/ab3800}
  {\bibfield  {journal} {\bibinfo  {journal} {The Astrophysical Journal}\
  }\textbf {\bibinfo {volume} {882}},\ \bibinfo {pages} {L24} (\bibinfo {year}
  {2019}{\natexlab{b}})}\BibitemShut {NoStop}%
\end{thebibliography}%

\end{document}